\documentstyle[aps,graphicx,epsfig]{revtex}\tighten
\draft

\def \D {\mbox{D}}

\def\be{\begin{equation}}
\def\ee{\end{equation}}
\def\bea{\begin{eqnarray}}
\def\eea{\end{eqnarray}}
\def\lb{\label}
\def\ct{\cite}
\def\r{\ref}

\def\gam{\gamma}
\def\d{\delta}
\def\eps{\epsilon}

\def\sig{\sigma}

\def\sigm{\sigma_{-}}
\def\sigc{\sigma_{\times}}
\def\np{n_{+}}
\def\nm{n_{-}}
\def\nc{n_{\times}}
\def\Sig{\Sigma}
\def\Sigp{\Sigma_{+}}
\def\Sigm{\Sigma_{-}}
\def\Sigc{\Sigma_{\times}}
\def\Np{N_{+}}
\def\Nm{N_{-}}
\def\Nc{N_{\times}}

\def\Om{\Omega}

\def\udot{\dot{u}}
\def\Udot{\dot{U}}
\def\ca{{\cal A}}

\def\cn{{\cal N}}

\def\D{\mbox{D}}
\def\p{{\bf e}}
\def\ptl{\partial}
\def\parb{\bbox{\partial}}
\def\e{{\rm e}}
\def\la{\langle}
\def\ra{\rangle}
\def\ti{\tilde}
\def\hsp5{\hspace{5mm}}
\newcommand{\sfrac}[2]{{\textstyle{#1\over#2}}}
\def\case#1/#2{\textstyle\frac{#1}{#2}}

\def\EEE{E_1{}^1}

\def\M{M^{\alpha\beta}}

\newcommand{\enl}{\\\hfill\rule{0pt}{0pt}}

\begin{document}

\title{Isotropic singularity in inhomogeneous brane cosmological models}

\author{A. A. Coley $^1$, Y. He$^1$, W. C. Lim$^{2}$} 
\address{$^1$ Department of Mathematics and Statistics,
Dalhousie University, Halifax, Nova Scotia}
\address{$^2$ Department of Applied Mathematics,  University of Waterloo, Waterloo, Ontario, Canada}

\address{E-mail: aac@mathstat.dal.ca and yanjing@mathstat.dal.ca}
\address{E-mail: wclim@math.uwaterloo.ca } \maketitle

\today
\begin{abstract}
We discuss the asymptotic dynamical evolution of spatially
inhomogeneous brane-world cosmological models close to the initial
singularity.  By introducing suitable
scale-invariant dependent variables and a suitable gauge, we write the evolution equations of the spatially inhomogeneous $G_{2}$ brane
cosmological models with one spatial degree of freedom as a system
of autonomous first-order partial differential equations. We study the 
system numerically, and we  find that there always 
exists an initial singularity, which is 
characterized by the fact that spatial derivatives are
dynamically negligible. More importantly, from the numerical analysis 
we conclude that there is an initial isotropic singularity  in all of these spatially inhomogeneous brane cosmologies for a range of parameter values which 
include the physically important cases of radiation and a scalar field 
source. The numerical results are supported by a qualitative dynamical analysis and a calculation of the past asymptotic decay rates. Although the analysis is local in nature, the numerics indicates that the singularity is isotropic for all relevant initial conditions.  Therefore this analysis,  and 
a preliminary investigation of general inhomogeneous ($G_0$) models, 
indicates that it is plausible that the  initial singularity is isotropic in spatially inhomogeneous brane-world cosmological models and consequently  that brane  cosmology naturally gives rise to a set of initial data that provide the 
conditions for inflation to subsequently take place.  
\end{abstract}


\section{Introduction}
String inspired theories in
which the matter fields are confined to a 3-dimensional
`brane-world' embedded in $1+3+d$ dimensions while the
gravitational field can also propagate in the $d$ extra dimensions 
\cite{rubakov} are currently of great interest. In particular, Randall and 
Sundrum~\cite{randall} have shown that for $d=1$, gravity can be localized 
on a single 3-brane at lower energies
even when the fifth dimension is infinite. The Randall-Sundrum type 
models are simple phenomenological models, which capture some of the 
essential features of the dimensional reduction of eleven-dimensional 
supergravity introduced by Ho$\check{\mbox{r}}$ava and 
Witten~\cite{Horava}. An elegant geometric
formulation and generalization of the Randall-Sundrum-type
brane-world models has been given~\cite{sms,Maartens}.  Recently there 
has been great interest in Randall-Sundrum-type brane-world cosmological
models  \cite{randall}, particularly in an attempt to understand the 
dynamics of the universe at early times. Brane-world models have a
different qualitative behaviour than their general-relativistic counterpart
\cite{BDL,sms}, especially at high energies when the energy density of 
the matter is larger than the  brane tension and  the behaviour  deviates 
significantly from the classical case.

The asymptotic dynamical evolution of spatially homogeneous brane-world 
cosmological models close to the initial singularity was studied in 
\cite{chaos}. It was found that an isotropic singularity \cite{GW85} is a 
past-attractor in all  orthogonal Bianchi models and is a local 
past-attractor in a class of inhomogeneous brane-world models (and 
consequently these models do not exhibit Mixmaster or chaotic-like 
behaviour  close to the initial singularity). However, the study of the 
behaviour of spatially homogeneous brane-worlds close to the initial 
singularity in the presence of  both local and nonlocal stresses 
indicates that for physically relevent values of the equation of state 
parameter there exist two local past attractors for these brane-worlds, 
one isotropic past attractor and one anisotropic past attractor; i.e.,  
in these brane-worlds the initial singularity can be locally either 
isotropic or anisotropic \cite{BH,hervik} (however, the anisotropic 
models appear to be unphysical and can likely be  ruled out). Therefore, it is plausible that typically the  initial singularity is isotropic in the 
brane-world scenario. Consequently, it was suggested  that brane 
cosmology naturally gives rise to a set of initial data that provide the 
conditions for inflation to subsequently take place, thereby solving  the 
initial conditions problem  and leading to a self--consistent and viable 
cosmology \cite{COLEY}.  We argued \cite{chaos} that it is plausible that
typically  the cosmological singularity is isotropic in spatially
inhomogeneous models.   We shall study this further here.

We shall consider the dynamics of a class of spatially inhomogeneous
cosmological models with one spatial degree of freedom in the
brane-world scenario. The $G_{2}$ cosmological models admit a
2-parameter Abelian isometry group acting transitively on
spacelike 2-surfaces. These models  admit one degree of freedom as
regards spatial inhomogeneity, and the resulting governing system
of evolution equations constitute a system of autonomous partial
differential equations in two independent variables. We follow the
formalism of \cite{elst} which utilizes area expansion normalized
scale-invariant dependent variables, and we use the timelike area
gauge to discuss the asymptotic evolution of the class of
orthogonally transitive $G_{2}$ cosmologies near the cosmological
initial singularity.  In this article we shall consider numerically the 
local dynamical behaviour of this class of spatially inhomogeneous models 
close to the singularity.

\subsection{Governing equations}

The field equations induced on the brane,  using the Gauss-Codazzi
equations, matching conditions and $Z_2$ symmetry, result in a
modification of the standard Einstein equations with the new terms
carrying bulk effects onto the brane \cite{sms,Maartens}:
\begin{equation}
G_{\mu\nu}=\kappa^2 T_{\mu\nu}+\widetilde{\kappa}^4S_{\mu\nu} -
{\cal E}_{\mu\nu}\,, \label{2}
\end{equation}
where
\begin{equation}
\kappa^2=8\pi/M_{\rm p}^2\,,
~~\lambda=6{\kappa^2\over\widetilde\kappa^4} \,\label{3}
\end{equation}

For {\em any} matter fields (scalar fields, perfect fluids,
kinetic gases, dissipative fluids, etc.), including a combination
of different fields, the general form of the brane energy-momentum
tensor can be covariantly given as
\begin{equation}
T_{\mu\nu}=\rho u_\mu
u_\nu+ph_{\mu\nu}+\pi_{\mu\nu}+2q_{(\mu}u_{\nu)}\,.
 \label{3''}
\end{equation}
The decomposition is irreducible for any chosen 4-velocity
$u^\mu$. Here $\rho$ and $p$ are the energy density and isotropic
pressure, and $h_{\mu\nu}=g_{\mu\nu}+u_\mu u_\nu$ projects
orthogonal to $u^\mu$. The energy flux obeys
$q_{\mu}=q_{\langle\mu\rangle}$, and the anisotropic stress obeys
$\pi_{\mu\nu}=\pi_{\langle\mu\nu\rangle}$, where angle brackets
denote the projected, symmetric and tracefree part:
\[
V_{\langle\mu\rangle}=h_\mu{}^\nu V_\nu\,,~~
W_{\langle\mu\nu\rangle}=\left[h_{(\mu}{}^\alpha h_{\nu)}{}^\beta-
{\textstyle{1\over3}}h^{\alpha\beta}h_{\mu\nu}\right]W_{\alpha\beta}\,.
\]
We shall primaily be interested in perfect fluid sources obeying the linear equation of state $p=(\gam-1)\,\rho$, where $\gam$ is constant with $0<\gam\leq 2$. The case $\gam =4/3$ corresponds to radiation. Scalar fields correspond to a stiff fluid $\gam =2$ close to the initial singularity.\\

The dynamical equations on the 3-brane differ from the general relativity
equations~\cite{sms,Maartens} in that there are nonlocal effects
from the free gravitational field in the bulk, transmitted via the
projection ${\cal E}_{\mu\nu}$ of the bulk Weyl tensor,  and  the local 
energy-momentum corrections, which are significant at very high energies 
and particularly close to the initial singularity. The matter fields 
contribute local quadratic energy-momentum corrections via the tensor 
$S_{\mu\nu}$, given by
\begin{equation}
S_{\mu\nu}={\textstyle{1\over12}}T_\alpha{}^\alpha T_{\mu\nu}
-{\textstyle{1\over4}}T_{\mu\alpha}T^\alpha{}_\nu+
{\textstyle{1\over24}}g_{\mu\nu} \left[3 T_{\alpha\beta}
T^{\alpha\beta}-\left(T_\alpha{}^\alpha\right)^2 \right]\,.
\label{3'}
\end{equation}

Equations (\ref{3''}) and (\ref{3'}) imply the irreducible
decomposition of \footnote{We take this opportunity to correct
equation (7) of~\cite{Maartens}.}
\begin{eqnarray}
S_{\mu\nu}&=&{\textstyle{1\over24}}\left[2\rho^2-3\pi_{\alpha\beta}
\pi^{\alpha\beta}\right]u_\mu u_\nu
+{\textstyle{1\over24}}\left[2\rho^2+4\rho p+\pi_{\alpha\beta}
\pi^{\alpha\beta}-4q_\alpha q^\alpha\right]h_{\mu\nu} \nonumber\\
&&{}-
{\textstyle{1\over12}}(\rho+3p)\pi_{\mu\nu}-{\textstyle{1\over4}}
\pi_{\alpha\langle\mu}
\pi_{\nu\rangle}{}^\alpha+{\textstyle{1\over4}}q_{\langle\mu}q_{\nu\rangle}+
{\textstyle{1\over3}}\rho q_{(\mu}u_{\nu)}- {\textstyle{1\over2}}
q^\alpha \pi_{\alpha(\mu}u_{\nu)} \,. \label{3'''}
\end{eqnarray}
The quadratic energy-momentum corrections to standard general relativity will be significant for $\widetilde{\kappa}^4\ti{\rho}^2 \gtrsim \kappa^2\ti{\rho}$ in the high-energy regime close to the singularity.

Nonlocal effects from the bulk can be irreducibly decomposed into
\begin{equation}
{\cal
E}_{\mu\nu}=-\left({\widetilde{\kappa}\over\kappa}\right)^4\left[{\cal
U}\left(u_\mu u_\nu+{\textstyle {1\over3}} h_{\mu\nu}\right)+{\cal
P}_{\mu\nu}+2{\cal Q}_{(\mu}u_{\nu)}\right]\,, \label{1}
\end{equation}
in terms of an effective nonlocal energy density on the brane,
${\cal U}$, arising from the free gravitational field in the bulk,
an effective nonlocal anisotropic stress on the brane, ${\cal
P}_{\mu\nu}$, and an effective nonlocal energy flux on the brane,
${\cal Q}_\mu$~\cite{Maartens}.

All of the bulk corrections may be consolidated into an effective
total energy density, pressure, anisotropic stress and energy
flux, as follows. The modified Einstein equations take the
standard Einstein form with a redefined energy-momentum tensor:
\begin{equation}
G_{\mu\nu}= T^{\rm tot}_{\mu\nu}\,, \label{8}
\end{equation}
where
\bea
T^{\rm tot}_{\mu\nu} & \equiv&
\kappa^2\,T_{\mu\nu}+{6\kappa^2\,\over \lambda}S_{\mu\nu}-
{\cal E}_{\mu\nu}\, \label{9}\\
&=&\rho^{\text{tot}} u_\mu
u_\nu+p^{\text{tot}}h_{\mu\nu}+\pi^{\text{tot}}_{\mu\nu}
+2q^{\text{tot}}_{(\mu}u_{\nu)}\,.
\eea
Then
\begin{eqnarray}
\rho^{\rm tot} &=& \kappa^2\,\rho+{6\kappa^2 \over
\lambda}\left[{1\over 24}\left(2\rho^2 -3
\pi_{\mu\nu}\pi^{\mu\nu}\right) + {1\over
\kappa^4}{\cal U}\right]\label{a}\\
p^{\rm tot} &=&\kappa^2\, p+ {6\kappa^2\over \lambda}\left[{1 \over 
24}\left(2\rho^2+4\rho p+
\pi_{\mu\nu}\pi^{\mu\nu}-4q_\mu q^\mu\right) +{1 \over 
3}{1\over\kappa^4}{\cal U}\right] \label{b}\\
 \pi^{\rm tot}_{\mu\nu} &=&
\kappa^2\,\pi_{\mu\nu}+ {6\,\kappa^2\over\lambda}\left[{1\over 
12}\left(-(\rho+3p)\pi_{\mu\nu}-3
\pi_{\alpha\langle\mu}\pi_{\nu\rangle}{}^\alpha+3q_{\langle\mu}q_
{\nu\rangle}\right) +{1\over \kappa^4}{\cal P}_{\mu\nu}\right]\label{c}\\
 q^{\rm tot}_\mu &=&\kappa^2\,q_\mu+ {6\,\kappa^2\over\lambda}\left[{1 
\over 12}\left(2\rho
q_\mu-3\pi_{\mu\nu}q^\nu\right)+ {1\over \,\kappa^4}{\cal Q}_\mu 
\right]\label{d}
\end{eqnarray}
(Note that $\widetilde{\kappa}^{4}/\kappa^6$ is dimensionless.)

As a consequence of the form of the bulk energy-momentum tensor and of 
$Z_2$ symmetry, it follows \cite{sms} that the brane energy-momentum 
tensor separately satisfies the conservation equations, i.e.,
\begin{equation}\label{12}
\nabla^\nu T_{\mu\nu}=0 \,.
\end{equation}
Consequently, the Bianchi identities on the brane imply that the projected 
Weyl tensor obeys the non-local constraint
\begin{equation}
\nabla^\mu{\cal E}_{\mu\nu}=\widetilde{\kappa}^4\nabla^\mu
S_{\mu\nu}\,. \label{5}
\end{equation}

The results of \cite{chaos} are incomplete in that a description
of the gravitational field in the bulk is not provided.
Unfortunately, the evolution of the anisotropic stress part is
{\em not} determined on the brane. The correction terms must be
consistently derived from the higher-dimensional equations. Since ${\cal P}_{\mu\nu}$ corresponds to gravitational
waves in higher-dimensions,  it is expected that the dynamics will
not be affected significantly at early times close to the
singularity (see \cite{waves} and later). Henceforth we shall effectively 
assume that
\begin{equation}
 {\cal P}_{\mu\nu}=0.\label{zeroPmunu}
\end{equation}
When ${\cal P}_{\mu\nu} =0$, the evolution of ${\cal E}_{\mu\nu}$ is fully determined \cite{sms}.
In the inhomogeneous cosmological models of interest here a non-zero
$\D_\mu{\ti{\rho}}$ acts as a source for ${\cal Q}_{\mu}$, and
hence ${\cal Q}_{\mu}=0$ is not consistent with an inhomogeneous
energy density, and we need to include a dynamical analysis of the
evolution of ${\cal Q}_{\mu}$. We shall make no further assumptions on 
the models and include all terms in the numerical analysis.

\subsection{Initial Singularity}

From the numerical analysis we shall find that the area expansion rate 
increases without bound ($\beta \rightarrow \infty$)  and the normalized 
frame variable \cite{elst} vanishes ($E_1{}^1 \rightarrow 0$) as 
logarithmic time $t \rightarrow -\infty$. Since $\beta \rightarrow 
\infty$ (and hence the Hubble rate diverges), there always 
exists an initial singularity as $t \rightarrow -\infty$. Thus the 
singularity is characterized by $E_1{}^1=0$, which allows both dynamical 
and numerical results to be obtained (see later).

In \cite{chaos} it was shown that the total energy density $\ti{\rho} \rightarrow \infty$
as $ t \rightarrow -\infty$. It then follows directly from the
conservation laws  that $\ti{\mu}_b \sim {\ti{\rho}_b}^2$
dominates as $ t \rightarrow -\infty$ and that all of the other
contributions to the brane energy density are negligible
dynamically as the singularity is approached. The fact that the
effective equation of state at high densities become ultra stiff,
so that the matter can dominate the shear dynamically, is a unique
feature of brane cosmology. Hence close to the singularity the matter 
contribution is effectively given by
\begin{eqnarray}
\ti{\rho}^{\rm tot} &=& {1\over 2\lambda}\ti{\rho}^2 \equiv
\ti{\mu}_b \\
\ti{p}^{\rm tot} &=& {1\over 2\lambda}(\ti{\rho}^2+2\ti{\rho}
\ti{p}) = (2\gamma -1)\ti{\mu}_b.
\end{eqnarray}
In addition, it follows from the conservation laws that $\Omega
\rightarrow 0$, $\Omega_{\cal U} \rightarrow 0$  and
$\Omega_\Lambda \rightarrow 0$ as the initial singularity is
approached. Hence, the models isotropize to the past. We shall study the 
generality of this result.

Models with an {\em isotropic initial singularity\/} \cite{GW85}
satisfy $\lim_{t \rightarrow -\infty}\Omega_b = 1$, $\lim_{t
\rightarrow -\infty}v = 0$, $\lim_{t \rightarrow
-\infty}\Sigma^{2} = 0$. Their evolution near the cosmological
initial singularity is approximated by the flat model corresponding to the `equilibrium point' ${\cal F}_b$, characterized by \footnote{All of the variables used here are defined later (e.g., see equations (\ref{dlframe})-(\ref{dlcurv1})).} 
\begin{equation} \Omega_b = 1; \, 0 = E_{1}{}^{1} = \Sigma_+ =\Sigma_- = \Sigma_{\times} = N_- = N_{\times} = v = Q_u = \Om=\Omega_u \ . \end{equation}
${\cal F}_b$  corresponds to a spatially homogeneous and isotropic
non-general-relativistic brane-world model (which is valid at very high
energies ($\ti{\rho} \gg \lambda$) as the initial singularity is
approached). Note that these solutions are
self-similar, and are  referred to as Bin\'etruy, Deffayet and
Langlois solutions \cite{BDL}\, or
Brane-Robertson-Walker models~\cite{COLEY}. It was shown that for all physically relevant values of
$\gamma$ ($\gamma \geq 1$), ${\cal F}_b$ is a source (or
past-attractor),  and hence the singularity is
isotropic,  in non-tilting  spatially homogeneous  brane-world models
\cite{chaos}. It was also shown that  ${\cal F}_b$ is a
local source or past-attractor in  the family of spatially
inhomogeneous `non-tilting' $G_2$ cosmological models for $\gamma >1$ \cite{chaos}. \\

In this paper we shall study the nature of the initial singularity in spatially inhomogeneous brane cosmological models. In particular, we shall study numerically the class of $G_2$ models. An analysis of the behaviour of spatially inhomogeneous solutions
to Einstein's equations near an initial singularity has been made in classical general relativity; in an investigation of a class of
Abelian $G_{2}$ spatially inhomogeneous models~\cite{inv} and a 
numerical investigation of a class of vacuum Gowdy $G_{2}$ 
cosmological spacetimes~\cite{WIB}, it was shown that the presence of the
inhomogeneity ceases to govern the dynamics asymptotically toward the
singularity.

\section{Brane $G_{2}$ cosmology}\lb{sec:BG2C}

We shall consider the class of  {\em $G_{2}$ cosmologies\/} with
two commuting Killing vector fields, which consequenly  admit one degree of  spatial freedom \cite{WE}. We shall follow the
approach of van Elst, Uggla and Wainwright \cite{elst}. The
evolution system of the EFE are partial differential
equations (PDE) in two independent variables. The {\em
orthonormal frame formalism\/} is utilized \cite{mac73,hveugg97} with the 
result that  (i) the governing equations are a first-order autonomous 
equation system, (ii) the  dependent variables are scale-invariant. 
In particular, we define scale-invariant dependent variables by normalisation with the area expansion rate of the $G_{2}$--orbits in order to obtain
the evolution equations as a system of PDE, ensuring the local existence,
uniqueness and stability of solutions to the Cauchy initial value
problem for $G_{2}$ cosmologies. Following \cite{elst} we 
assume that the Abelian $G_{2}$ isometry group acts {\em
orthogonally transitively\/} on spacelike 2-surfaces, and
introduce a group-invariant orthonormal frame $\{\,\p_{a}\,\}$,
with $\p_{2}$ and $\p_{3}$ tangent to the $G_{2}$--orbits. The frame vector field $\p_{0}$,  which defines a {\em
future-directed timelike reference congruence\/}, 
is orthogonal to the $G_{2}$--orbits and it is hypersurface
orthogonal and hence is orthogonal to a locally defined family of
spacelike 3-surfaces $t=\mbox{const}$. We then introduce
a set of symmetry-adapted local coordinates $\{\,t, \,x, \,y,
\,z\,\}$ 
\be
\lb{framecompos}
\p_{0} = N^{-1}\,\ptl_{t} \ , \hsp5
\p_{1} = e_{1}{}^{1}\,\ptl_{x} \ , \hsp5
\p_{A} = e_{A}{}^{B}\,\ptl_{x^{B}} \ , \hsp5
A, \,B = 2, \,3 \ ,
\ee
where the coefficients are functions of the independent variables
$t$ and $x$ only. The only non-zero {\em frame variables\/} are
thus given by
\be
N \ , \hsp5 e_{1}{}^{1} \ , \hsp5 e_{A}{}^{B} \ ,
\ee
which yield the following non-zero {\em connection variables\/} \ct{hveugg97}:
\be \alpha, \,\beta, \,a_{1}, \,\np, \,\sigm, \,\nc, \,\sigc,
\,\nm, \,\udot_{1}, \,\Om_{1}. \ee The variables $\alpha$, $\beta$, $\sigm$ and
$\sigc$ are related to the Hubble volume expansion rate $H$ and
the shear rate $\sig_{\alpha\beta}$ of the timelike reference
congruence $\p_{0}$; in particular, $ \Theta := 3H = \alpha + 2\beta$. 
The variables $a_{1}$, $\np$, $\nc$ and $\nm$ describe the
non-zero components of the purely spatial commutation functions
$a^{\alpha}$ and $n_{\alpha\beta}$ \cite{WE}. Finally, the
variable $\udot_{1}$ is the acceleration of the timelike reference
congruence $\p_{0}$, while $\Om_{1}$ represents the rotational
freedom of the spatial frame $\{\,\p_{\alpha}\,\}$ in the $({\bf
e}_{2},{\bf e}_{3})$--plane. Setting $\Om_{1}$ to zero corresponds
to the choice of a Fermi-propagated orthonormal frame
$\{\,\p_{a}\,\}$.  Within the present
framework the dependent variables
\be
\{\,N, \,\udot_{1}, \,\Om_{1}\,\}
\ee
enter the evolution system as freely prescribable {\em gauge
source functions\/}.

Since the $G_{2}$ isometry group acts orthogonally transitively,
the 4-velocity vector field $\ti{\bf u}$ of the perfect fluid is
orthogonal to the $G_{2}$--orbits, and hence has the form
\be
\lb{fluid4vel}
\ti{\bf u} = \Gamma\,(\p_{0}+v\,\p_{1}) \ ,
\ee
where $\Gamma \equiv (1-v^{2})^{-1/2}$.

We assume that the ordinary matter is a perfect fluid with equation of state
\be
	p_{fl} = (\gamma-1) \rho_{fl}\ .
\ee
In a tilted frame, we have
\be
	T_{\mu\nu} = \rho u_\mu u_\nu + p h_{\mu\nu} + \pi_{\mu\nu}
			+ 2 q_{(\mu} u_{\nu)}\ ,
\ee
where
\be
\label{matter}
	\rho = \frac{G_+}{1-v^2} \rho_{fl},\quad
	p = \frac{1}{3} \frac{3(\gamma-1)(1-v^2) + \gamma v^2}{G_+} 
		\rho,\quad
	q_\alpha = \frac{\gamma \rho}{G_+} v_\alpha,\quad
	\pi_{\alpha\beta} = \frac{\gamma \rho}{G_+} v_{\la\alpha} 
	v_{\beta\ra}
\ee
and $G_+ = 1+(\gamma-1)v^2$.
The basic variables that we use are $\rho$ and $v^\alpha$.

The quadratic correction matter tensor $S_{\alpha\beta}$ is given by

\be
	S_{\mu\nu} = \rho^b u_\mu u_\nu + p^b h_{\mu\nu} + \pi^b{}_{\mu\nu}
                        + 2 q^b{}_{(\mu} u_{\nu)}\ ,
\ee
where

\bea
	\rho^b &=& \frac{1}{24} (2\rho^2-3 \pi_{\alpha\beta} 
			\pi^{\alpha\beta})
\\
		&= &\frac{1}{12} \frac{(1-v^2)}{G_+{}^2} 
			[ 1+(2\gamma-1)v^2 ] \rho^2\ ,
\label{rho_b_rho}
\\
	p^b &=& \frac{1}{24} ( 2\rho^2 + 4\rho p + \pi_{\alpha\beta} 
			\pi^{\alpha\beta} - 4 q_\alpha q^\alpha)
\\
		&=& \frac{1}{3} \frac{3(2\gamma-1)(1-v^2) + 2\gamma 
			v^2}{1+(2\gamma-1)v^2} \rho^b\ ,
\\
	q^b_\alpha &=& \frac{1}{12} ( 2\rho q_\alpha - 3 q^\beta 
			\pi_{\beta\alpha})
\\
		&=& \frac{2\gamma \rho^b}{1+(2\gamma-1)v^2} v_\alpha\ ,
\\
	\pi^b_{\alpha\beta} &=& \frac{1}{12} [ -(\rho+3p) \pi_{\alpha\beta}
			-3 \pi_{\gamma\la\alpha}\pi_{\beta\ra}{}^\gamma 
			+ 3 q_{\la\alpha} q_{\beta\ra} ]
\\
	&=& \frac{2\gamma \rho^b}{1+(2\gamma-1)v^2} v_{\la\alpha}v_{\beta\ra}\ .
\eea     
Comparing with (\ref{matter}), we see that the quadratic matter is effectively a  perfect fluid with equation of state
\be
        p^b{}_{fl} = (2\gamma-1) {\rho^b}_{fl}\ .
\ee

From equation (\ref{1}), the bulk matter ${\cal E}_{\mu\nu}$ is given by
\be
{\cal
E}_{\mu\nu} = -{6\kappa^2\over\lambda}(\rho^u u_\mu u_\nu + p^u h_{\mu\nu} 
			+ \pi^u{}_{\mu\nu} + 2 q^u{}_{(\mu} u_{\nu)})
\ee
where
\bea
	\rho^u &=& {1\over\kappa^4}{\cal U}=\frac{ 1 + \frac{1}{3} v^2}{1-v^2} {\rho^u}_{fl}
\\
	p^u &=&{1\over 3}{1\over\kappa^4}{\cal U}= \frac{1}{3} \rho^u
\\
	q^u{}_\mu &=&{1\over\kappa^4}{\cal Q_\mu}= \frac{\frac{4}{3} \rho^u}{1+\frac{1}{3}v^2} v_\mu
			+ \frac{1}{\sqrt{1-v^2}} (q^u{}_{fl})_\mu
\\
	\pi^u{}_{\mu\nu} &=&{1\over\kappa^4}{\cal P_{\mu\nu}}= \frac{\frac{4}{3} \rho^u}{1+\frac{1}{3}v^2} 
			v_{\la\mu}v_{\nu\ra}
			+ (\pi^u{}_{fl})_{\mu\nu}\ .
\eea
Comparing with (\ref{matter}), we see that the bulk matter is similar to a
perfect fluid with equation of state
\be
        p^u{}_{fl} = \frac{1}{3} {\rho^u}_{fl}\ ,
\ee
but with non-zero $(q^u{}_{fl})_\mu$ and 
$(\pi^u{}_{fl})_{\mu\nu}$. We shall use $\rho^u$ and $q^u{}_\mu$ as the 
basic variables, and set $(\pi^u{}_{fl})_{\mu\nu}=0$ in this paper (see~(\ref{zeroPmunu})).\\

Since 
\[q^u{}_{\mu}=q^u\delta_{\mu}{}^1, \hsp5 v_{\mu}=v\delta_{\mu}{}^1, \hsp5 (\pi^u{}_{fl})_{\mu\nu}=0,\]
we can define  $q$, $q_b$, $q_u$, $\pi$, $\pi_b$ and $\pi_u$ as follows:
\[ (q_1,\, q_2, \, q_3)= (q,\,0,\,0),  \hsp5 (q^b{}_1,\, q^b{}_2, \, q^b{}_3)= (q_b,\,0,\,0),  \hsp5 (q^u{}_1,\, q^u{}_2, \, q^u{}_3)= (q_u,\,0,\,0)\]
and 
\[  \pi_{\mu\nu}=0  \hsp5 \text{when} \hsp5 \mu \not= \nu. \hsp5 \hsp5   \pi_{11}=\pi, \hsp5 \pi_{22}=\pi_{33}= -\frac{\pi_{11}}{2}= -\frac{\pi}{2} \]
\[  \pi^b{}_{\mu\nu}=0  \hsp5 \text{when} \hsp5 \mu \not= \nu. \hsp5 \hsp5   \pi^b{}_{11}=\pi_b, \hsp5 \pi^b{}_{22}=\pi^b{}_{33}= -\frac{\pi^b{}_{11}}{2}= -\frac{\pi_b}{2} \]
\[  \pi^u{}_{\mu\nu}=0  \hsp5 \text{when} \hsp5 \mu \not= \nu. \hsp5 \hsp5   \pi^u{}_{11}=\pi_u, \hsp5 \pi^u{}_{22}=\pi^u{}_{33}= -\frac{\pi^u{}_{11}}{2}= -\frac{\pi_u}{2} \]

The orthonormal frame version of the EFE, matter equations and
non-local equations, when specialised to the orthogonally
transitive Abelian $G_{2}$ case with the dependent variables
presented above~\ct{elst}, takes the following form: \enl

\noindent
{\bf Einstein field equations and Jacobi identities} \nopagebreak

\noindent
{\em Evolution equations\/}:
\bea
\lb{alphadot}
N^{-1}\,\ptl_{t}\alpha & = & -\,\alpha^{2} + \beta^{2}
- 3\,(\sigm^{2}-\nc^{2}+\sigc^{2}-\nm^{2})
- a_{1}^{2} + (e_{1}{}^{1}\,\ptl_{x}+\udot_{1})\,\udot_{1} \nonumber \\
& & \hsp5 - \sfrac{1}{2}(\rho^{\text{tot}} + p^{\text{tot}}) +\pi^{\text{tot}}
\\
\lb{betadot}
N^{-1}\,\ptl_{t}\beta & = & -\,\sfrac{3}{2}\,\beta^{2}
- \sfrac{3}{2}\,(\sigm^{2}+\nc^{2}+\sigc^{2}+\nm^{2})
- \sfrac{1}{2}\,(2\udot_{1}-a_{1})\,a_{1} \nonumber \\
& & \hsp5 - \sfrac{1}{2}(p^{\text{tot}} + \pi^{\text{tot}})\\
\lb{a1dot}
N^{-1}\,\ptl_{t}a_{1} & = & -\,\beta\,(\udot_{1}+a_{1})
- 3\,(\nc\,\sigm-\nm\,\sigc)
- \sfrac{1}{2}q^{\text{tot}}\\
\lb{npdot}
N^{-1}\,\ptl_{t}\np & = & -\,\alpha\,\np
+ 6\,(\sigm\,\nm+\sigc\,\nc)
- (e_{1}{}^{1}\,\ptl_{x}+\udot_{1})\,\Om_{1} \\
\lb{sigmdot}
N^{-1}\,\ptl_{t}\sigm + e_{1}{}^{1}\,\ptl_{x}\nc
& = & -\,(\alpha+2\beta)\,\sigm
- 2\,\np\,\nm - (\udot_{1}-2a_{1})\,\nc - 2\,\Om_{1}\,\sigc \\
\lb{ncdot}
N^{-1}\,\ptl_{t}\nc + e_{1}{}^{1}\,\ptl_{x}\sigm
& = & -\,\alpha\,\nc + 2\,\sigc\,\np - \udot_{1}\,\sigm
+ 2\,\Om_{1}\,\nm \\
\lb{sigcdot}
N^{-1}\,\ptl_{t}\sigc - e_{1}{}^{1}\,\ptl_{x}\nm
& = &  -\,(\alpha+2\beta)\,\sigc
- 2\,\np\,\nc + (\udot_{1}-2a_{1})\,\nm + 2\,\Om_{1}\,\sigm \\
\lb{nmdot}
N^{-1}\,\ptl_{t}\nm - e_{1}{}^{1}\,\ptl_{x}\sigc
& = & -\,\alpha\,\nm + 2\,\sigm\,\np + \udot_{1}\,\sigc
- 2\,\Om_{1}\,\nc \ .
\eea

\noindent
{\em Constraint equations\/}:
\bea \lb{gauss} 0 & = &  \ 2\,(2\,e_{1}{}^{1}\,\ptl_{x}
-3\,a_{1})\,a_{1} - 6\,(\nc^{2}+\nm^{2}) +
2\,(2\alpha+\beta)\,\beta
- 6\,(\sigm^{2}+\sigc^{2}) - \,2\rho^{\text{tot}} \\
\lb{codac} 0 & = & \ e_{1}{}^{1}\,\ptl_{x}\beta +
a_{1}\,(\alpha-\beta) - 3\,(\nc\,\sigm-\nm\,\sigc) - \sfrac{1}{2}q^{\text{tot}}. \eea

\noindent {\bf Bianchi identities (conservation equations)}
\nopagebreak

\noindent
{\em Evolution equations\/}:
\bea \lb{mudot} \frac{f_{1}}{\rho}\,(N^{-1}\,\ptl_{t} +
\frac{\gam}{G_{+}}\,v\,e_{1}{}^{1}\,\ptl_{x})\,\rho + 
f_{2}\,e_{1}{}^{1}\,\ptl_{x}v & = &
-\,\frac{\gam}{G_{+}}\,f_{1}\,[\ \alpha\,(1+v^{2})
+ 2\beta + 2\,(\udot_{1}-a_{1})\,v\ ] \\
\lb{vdot} \frac{f_{2}}{f_{1}}\,\rho\,(N^{-1}\,\ptl_{t} -
\frac{f_{3}}{G_{+}G_{-}}\,v\,e_{1}{}^{1}\,\ptl_{x})\,v +
f_{2}\,e_{1}{}^{1}\,\ptl_{x}\rho & = &
-\,\frac{f_{2}}{f_{1}G_{-}}\,\rho\,(1-v^{2})\,
[\ (2-\gam)\,\alpha\,v - 2\,(\gam-1)\,\beta\,v \nonumber \\
& & \hspace{20mm} + \ G_{-}\,\udot_{1}
+ 2\,(\gam-1)\,a_{1}\,v^{2}\ ] \ ,
\eea
where
\be
\lb{fdef}
f_{1} := \frac{(\gam-1)}{\gam G_{-}}\,(1-v^{2})^{2} \ , \hsp5
f_{2} := \frac{(\gam-1)}{G_{+}^{2}}\,(1-v^{2})^{2} \ , \hsp5
\ee
\be
\lb{gpmdef}
G_\pm := 1\pm (\gam-1)\,v^2 \ , \hsp5  f_{3} := (3\gam-4)-(\gam-1)\,(4-\gam)\,v^{2} \ .
\ee
From equation (\ref{5}) we obtain:\\
\noindent {\bf Non-local conservation equations} \nopagebreak

\noindent {\em Evolution equation\/}:
\bea
	N^{-1}\,\ptl_{t}\rho_u +e_1{}^1\ptl_{x}q_u &=&
        -\frac{4}{3+v^2}\left[(1+v^2)\alpha+2\beta\right]
	\rho_u-2(\dot{u}_1-a_1)q_u
	+vY
\\
	N^{-1}\,\ptl_{t}q_u + e_1{}^1\ptl_{x} (p_u+\pi_u) &=&
	-\frac{4}{3+v^2}\left[(1+v^2)\dot{u}_1-2v^2a_1\right]
	\rho_u-2(\alpha+\beta)q_u
	+Y
\eea
where
\bea
	\ptl_{x}(p_u+\pi_u) &=&
	\frac{1+3v^2}{3+v^2}\ptl_x \rho_u +\frac{16v}{(3+v^2)^2}\rho_u 
	\ptl_x v
\\
	Y &\equiv& \frac{\gam^2 v(1-v^2)\rho^2}{6 G_+{}^2G_-}
	\left[ (1-v^2)\alpha + 2\beta - 2 v a_1 \right]
\nonumber\\&&
	-\frac{\gam(1-v^2)^2\rho}{6G_+{}^2G_-} e_1{}^1\ptl_{x} \rho
	+\frac{\gam^2v(1-v^2)(3+(\gam-1)v^2)\rho^2}{6G_+{}^3G_-}
	 e_1{}^1\ptl_{x} v\ ,
\eea

\noindent and
\bea
\rho^{\text{tot}}&=& \kappa^2\rho + {6\kappa^2 \over \lambda}\rho_b +{6\kappa^2 \over \lambda}\rho_u \nonumber\\
&=&\kappa^2 \rho +{6\kappa^2 \over \lambda}\frac{(1-v^2)}{12\,G_+{}^2}\left(1+(2\,\gam-1)v^2\right)\rho^2  +{6\kappa^2 \over \lambda}\rho_u \\
p^{\text{tot}}&=& \kappa^2 p +{6\kappa^2 \over \lambda} p_b +{6\kappa^2 \over \lambda}p_u \nonumber\\
 & =& \frac{1}{3}\frac{3(\gam-1)(1-v^2)+\gam\,v^2}{G_+}\kappa^2\rho +
 {6\kappa^2 \over \lambda}\frac{1}{3}\frac{3(2\,\gam-1)(1-v^2)+2\,\gam\,v^2}{12\,G_+{}^2}(1-v^2)\rho^2 +{6\kappa^2 \over \lambda}\frac{1}{3}\rho_u      \\
q^{\text{tot}}&=& \kappa^2q +{6\kappa^2 \over \lambda} q_b +{6\kappa^2 \over \lambda}q_u \nonumber   \\
& =& \frac{\gam\,v}{G_+}\kappa^2\rho + {6\kappa^2 \over \lambda}\frac{\gam\,v\,(1-v^2)}{6\,G_+{}^2}\rho^2  + {6\kappa^2 \over \lambda}q_u   \\
\pi^{\text{tot}}&=& \kappa^2\pi + {6\kappa^2 \over \lambda}\pi_b +{6\kappa^2 \over \lambda}\pi_u \nonumber\\
& =& \frac{2}{3}\frac{\gam\,v^2}{G_+}\kappa^2\rho + {6\kappa^2 \over \lambda}\frac{2}{3}\frac{\gam\,v^2\,(1-v^2)}{6\,G_+{}^2}\rho^2  +{6\kappa^2 \over \lambda} \frac{2}{3}\frac{4\,v^2}{(3+v^2)}\rho_u
\eea
Finally, we have:

\noindent {\em Gauge fixing condition\/}:
\be \lb{gaugefix} 0 =  N^{-1}\,e_{1}{}^{1}\,\ptl_{x}N - \udot_{1}
\ . \ee The frame variables $e_{A}{}^{B}$ decouple from the
remaining equations and we need not consider them further.

\subsection{Scale-invariant reduced equation system}
\lb{subsec:dimlesseq}
 We
introduce $\beta$-normalised frame, connection and curvature
variables as follows \cite{elst}\footnote{Expressions for the area expansion rate and the scale-invariant dependent variables in terms of the line element, written in the separable area gauge, are given in  \cite{elst}.}:
\bea
\lb{dlframe}
(\,\cn^{-1}, \,E_{1}{}^{1}\,)
& := & (\,N^{-1}, \,e_{1}{}^{1}\,)/\beta \\
(\,\Udot, \,A, \,(1-3\Sigp), \,\Sigm, \,\Nc, \,\Sigc, \,\Nm,
\,\Np, \,R\,)
& := & (\,\udot_{1}, \,a_{1}, \,\alpha, \,\sigm, \,\nc,
\,\sigc, \,\nm, \,\np, \,\Om_{1}\,)/\beta \\
\lb{dlcurv1} (\,\Om, \,\Om_u, \,Q_u \,) & := & (\,\kappa^2\,\rho, \, 
{6\kappa^2 \over \lambda}\rho_u, 
\,{6\kappa^2 \over \lambda}q_u\,)/(3\,\beta^2)=(\,\kappa^2\,\rho, \, 
\frac{6 {\cal U}}{\lambda\kappa^2},
\, \frac{6{\cal Q}}{\lambda\kappa^2}\,)/(3\,\beta^2)\\
  B & := & \sqrt{\frac{6}{\lambda\kappa^2}}\sqrt{3}\beta
\eea
 where $\beta$ is the area
expansion rate of the $G_{2}$--orbits. The new {\em dimensionless
dependent variables\/}  are invariant under arbitrary scale
transformations, and are linked to the $H$-normalised variables
through the relation $H = (1-\Sigp)\,\beta$ \cite{chaos,WE}. Note that in the units we have chosen the matter
variable $v$ {\em is\/} already dimensionless.

In order to write the dimensional equation system in terms of scale-invariant dependent variables
(\r{dlframe}) -- (\r{dlcurv1}) it is necessary to introduce the
time and space rates of change of the normalisation factor
$\beta$. We use the
evolution equation (\r{betadot}) and the 
constraint equation (\r{codac}) to define  $q$ and $r$ in terms of the remaining scale-invariant dependent
variables:

\bea
\lb{betaq}
\cn^{-1}\,\ptl_{t}B & = & -\,(q+1)\,B \\
\lb{betar} 0 & = & (E_{1}{}^{1}\,\ptl_{x}+r)\,B \ , 
\eea
the expressions (\r{hdecel}) and (\r{codac1}) (below) for $q$ and $r$ are {\em purely algebraical\/}, and are referred to  as {\em the defining equations} for $q$
and $r$\/; $q$ and $r$ play the r\^{o}le of an
``area deceleration parameter'' and a r\^{o}le analogous to a
``Hubble spatial gradient''. Using equations (\r{betaq}) and
(\r{betar}), the definitions (\r{dlframe}) -- (\r{dlcurv1}) and equation (13) in \cite{elst}, it is straightforward to transform the dimensional equation system to a $\beta$-normalised dimensionless form.  

\noindent {\bf Scale-invariant equation system} 

\noindent
{\em Evolution system\/}:
\bea
\cn^{-1}\,\ptl_{t}B
& = & -(q+1)\,B \\
\lb{dle11dot}
\cn^{-1}\,\ptl_{t}E_{1}{}^{1}
& = & (q+3\Sigp)\,E_{1}{}^{1} \\
\lb{dlsigpdot}
\cn^{-1}\,\ptl_{t}\Sigp
& = & (q+3\Sigp - 2)\,\Sigp-2\,(\Nm^2 + \Nc ^2)-\sfrac{1}{3}\,E_1{}^1\,\ptl_{x}r - \sfrac{3}{2}\,\Pi_{\text{total}}\\
\lb{dladot}
\cn^{-1}\,\ptl_{t}A
& = & (q+3\Sigp)\,A + (r-\Udot) \\
\lb{dlnpdot}
\cn^{-1}\,\ptl_{t}\Np
& = & (q+3\Sigp)\,\Np + 6\,(\Sigm\,\Nm+\Sigc\,\Nc)
- (E_1{}^1\,\ptl_{x}-r+\Udot)\,R \\
\lb{dlsigmdot}
\cn^{-1}\,\ptl_{t}\Sigm + E_{1}{}^{1}\,\ptl_{x}\Nc
& = & (q+3\Sigp-2)\,\Sigm - 2\,\Np\,\Nm
+ (r-\Udot+2A)\,\Nc - 2\,R\,\Sigc \\
\lb{dlncdot}
\cn^{-1}\,\ptl_{t}\Nc + E_{1}{}^{1}\,\ptl_{x}\Sigm
& = & (q+3\Sigp)\,\Nc + 2\,\Sigc\,\Np
+ (r-\Udot)\,\Sigm + 2\,R\,\Nm \\
\lb{dlsigcdot}
\cn^{-1}\,\ptl_{t}\Sigc - E_{1}{}^{1}\,\ptl_{x}\Nm
& = & (q+3\Sigp-2)\,\Sigc - 2\,\Np\,\Nc
- (r-\Udot+2A)\,\Nm + 2\,R\,\Sigm \\
\lb{dlnmdot}
\cn^{-1}\,\ptl_{t}\Nm - E_{1}{}^{1}\,\ptl_{x}\Sigc
& = & (q+3\Sigp)\,\Nm + 2\,\Sigm\,\Np
- (r-\Udot)\,\Sigc - 2\,R\,\Nc \ .
\eea
\bea \lb{dlmudot} \frac{f_{1}}{\Om}\,(\cn^{-1}\,\ptl_{t} +
\frac{\gam}{G_{+}}\,v\,E_{1}{}^{1}\,\ptl_{x})\,\Om +
\,f_{2}\,E_{1}{}^{1}\,\ptl_{x}v & = &
2\,\frac{\gam\,f_{1}}{G_{+}}\,[\ \frac{G_{+}}{\gam}\,(q+1)
-\sfrac{1}{2} \,(1-3\Sigp)\,(1+v^{2}) \nonumber \\
& & \hspace{25mm} - \,1 + (r-\Udot+A)\,v\ ] \\
\lb{dlvdot} \frac{f_{2}}{f_{1}}\,\Om\,(\cn^{-1}\,\ptl_{t} -
\frac{f_{3}}{G_{+}G_{-}}\,v\,E_{1}{}^{1}\,\ptl_{x})\,v +
\,f_{2}\,E_{1}{}^{1}\,\ptl_{x}\Om & = &
2\,\frac{f_{2}}{f_{1}G_{-}}\,\Om\,(1-v^{2})\,
[\ \frac{(\gam-1)}{\gam}\,(1-v^{2})\,r  \nonumber\\
& & \hspace{15mm} - \ \sfrac{1}{2}\,(2-\gam)\,(1-3\Sigp)\,v
\nonumber \\
& & \hspace{15mm} + \ (\gam-1)\,(1-A\,v)\,v -
\sfrac{1}{2}\,G_{-}\,\Udot\ ] \eea
\bea
        \cn^{-1}\,\ptl_t \Omega_u + \EEE \ptl_x Q_u
        &=& \left[2(q-1)+12\frac{1+v^2}{3+v^2}\Sigp\right]\Omega_u
                -2(\dot{U}-r-A)Q_u +vX
\\
        \cn^{-1}\,\ptl_t Q_u + \EEE \ptl_x (P_u + \Pi_u)
        &=& 2(q+3\Sigp-1)Q_u
        -\frac{2}{3+v^2}\left[2(1+v^2)(\dot{U}-r)-4v^2A
        +(1-v^2)r \right]\Omega_u +X
\eea
where
\bea
        \ptl_x (P_u + \Pi_u) &=&
        \frac{1+3v^2}{3+v^2} \, \ptl_x \Omega_u
                + \frac{16 v}{(3+v^2)^2}\Omega_u \, \ptl_x v
\\
        X &\equiv& \frac{\gamma(1-v^2) B^2 \Omega^2}{6 {G_+}^2 G_-}
        \Big[ \gamma v (1-v^2)(1-3\Sigp) +2\gamma v -2\gam v^2A
         +  2(1-v^2) r \Big]
\nonumber\\
        &&\quad - \frac{\gamma (1-v^2)^2}{6{G_+}^2G_-} B^2 \Omega
        \,\EEE \ptl_x \Omega
        + \frac{\gamma^2v(1-v^2)(3+(\gamma-1)v^2)}{6{G_+}^3G_-}
                B^2 \Omega^2 \,\EEE \ptl_x v\ .
\eea
{\em Constraint equations\/}:
\bea 
\lb{fried} \sfrac{2}{3}\,E_{1}{}^{1}\,\ptl_{x}\,A  &=&  2\Sigp - 1
+ \Sigm^{2} + \Sigc^{2} +
\Nc^{2} + \Nm^{2}  + \,\sfrac{2}{3}\,r\,A + A^{2}+\Om_{\text{total}}\\
 0 & = & r +\,3\,A\,\Sigp + 3\,(\Nc\,\Sigm-\Nm\,\Sigc) +
\,\sfrac{3}{2}\,Q_{\text{total}} 
 \eea
\noindent {\em Defining equations for $q$ and $r$\/}:
\bea
\lb{hdecel}
q & := & \sfrac{1}{2} + \sfrac{1}{2}\,(2\Udot-A)\,A
+ \sfrac{3}{2}\,(\Sigm^{2}+\Nc^{2}+\Sigc^{2}+\Nm^{2})
+ \sfrac{3}{2}(\,P_{\text{total}} +\,\Pi_{\text{total}})\\
\lb{codac1} r & := &-\frac{ E_1{}^1\,\ptl_{x}B}{B}
\eea
where 
\bea
\Om_{\text{total}}&=& \Om + \Om_b +\Om_u \nonumber\\
\lb{Om_tot}
&=& \Om 
+\frac{(1-v^2)}{12\,G_+{}^2}\left(1+(2\,\gam-1)v^2\right)\Om^2\,B^2  
+\Om_u \\
P_{\text{total}}&=& P + P_b +P_u \nonumber\\
 & =& \frac{3(\gam-1)(1-v^2)+\gam\,v^2}{3\,G_+}\Om +
 \frac{3(2\,\gam-1)(1-v^2)+2\,\gam\,v^2}{36\,G_+{}^2}(1-v^2)\Om^2\,B^2 
+\frac{1}{3}\Om_u      \\
Q_{\text{total}}&=& Q + Q_b +Q_u \nonumber   \\
& =& \frac{\gam\,v}{G_+}\Om + 
\frac{\gam\,v\,(1-v^2)}{6\,G_+{}^2}\Om^2\,B^2  + Q_u   \\
\Pi_{\text{total}}&=& \Pi + \Pi_b +\Pi_u \nonumber\\
\lb{Pi_tot}
& =& \frac{2}{3}\frac{\gam\,v^2}{G_+}\Om + 
\frac{\gam\,v^2\,(1-v^2)}{9\,G_+{}^2}\Om^2\,B^2  + 
\frac{8\,v^2}{3(3+v^2)}\Om_u
\eea
define the various physical quantities. In particular, we note that
\be \label{defomb} 
\Om_b \equiv 
\frac{(1-v^2)}{12\,G_+{}^2}\left(1+(2\,\gam-1)v^2\right)\Om^2\,B^2\,.
\ee
{\em Gauge fixing condition\/}:
\be \lb{dlgsfr} 0 = \cn^{-1}\,E_{1}{}^{1}\,\ptl_{x}\cn + (r-\Udot)\ . 
\ee
\enl In the above, $f_{1}$, $f_{2}$, $f_{3}$ and $G_{\pm}$ are
defined by equations. (\r{fdef}) and (\r{gpmdef}), respectively.

\subsection{Gauge choice} \lb{sec:gauge}

The scale-invariant equation system in
subsection~\r{subsec:dimlesseq} contains evolution equations for
the dependent variables
\be \{\,E_{1}{}^{1}, \,\Sigp, \,A, \,\Np, \,\Sigm, \,\Nc, \,\Sigc,
\,\Nm, \,\Om_b, \,v,\, \,Q\,\} \ , \ee
but not for the gauge source functions
\be
\{\,\cn, \,\Udot, \,R\,\} \ ,
\ee
which are arbitrarily prescribable real-valued functions of the
independent variables $t$ and $x$, and thus does not uniquely
determine the evolution of the $G_{2}$ cosmologies. The reason for
this deficiency is that the orthonormal frame $\{\,\p_{a}\,\}$ and
the local coordinates $\{\,t, \,x\,\}$ have not been specified uniquely. The remaining {\em gauge freedom\/} consists of a choice timelike reference congruence
$\p_{0}$ and of local time and space coordinates $t$ and $x$ (the
{\em temporal gauge freedom\/}), and a choice of spatial frame
vector fields $\p_{2}$ and $\p_{3}$ (the {\em spatial gauge
freedom\/}).\\

We shall fix the spatial gauge by requiring \be \lb{sgfix} \Np =
\sqrt{3}\,\Nm \ , \hsp5 R = -\,\sqrt{3}\,\Sigc \ ,\ee which is
preserved under evolution and under a boost. With this choice the
evolution equation (\r{dlnpdot}) becomes identical to equation.
(\r{dlnmdot}), and thus can be omitted from the full
scale-invariant equation system.

We fix the temporal gauge by adapting the evolution of the gauge
source function $\Udot$. The {\em separable area gauge} is
determined by imposing the condition
\be \lb{areagc} 0 = (r-\Udot) \ , \ee
which determines $\Udot$ algebraically through equation. (\r{codac1}).
There is thus no need to determine an evolution equation for
$\Udot$. It follows immediately from the gauge fixing condition
(\r{dlgsfr}) that $\cn = f(t)$. We now use the
$t$-reparametrisation to set $f(t) = \cn_{0}$, a constant, which
we choose to be unity, i.e.,
\be \lb{areagn} \cn = \cn_{0} := 1\ . \ee
Therefore, $t$ is effectively a logarithmic proper time, and the initial singularity occurs for $t\rightarrow -\infty$.\\

The {\em area density\/} $\ca$ of the $G_{2}$--orbits plays a
prominent r\^{o}le for $G_{2}$ cosmologies. Expressed in terms of
the coordinate components of the frame vector fields $\p_{A}$
tangent to the $G_{2}$--orbits this becomes
\be \ca^{-1} = e_{2}{}^{2}\,e_{3}{}^{3} - e_{2}{}^{3}\,e_{3}{}^{2}
\ . \ee
In terms of our scale-invariant dependent
variables, the area density $\ca$ of the $G_{2}$--orbits satisfies
the relations \cite{elst}
\be \lb{dad1} \ca^{-1}\,\cn^{-1}\,\ptl_t\ca = 2 \ , \hsp5
\ca^{-1}\,E_{1}{}^{1}\,\ptl_{x}\ca = -\,2A \ . \ee
Combining the two, the magnitude of the spacetime gradient
$\nabla_{a}\ca$ is
\be \lb{agrad} (\nabla_{a}\ca)\,(\nabla^{a}\ca) =
-\,4\beta^{2}\,(1-A^{2})\,\ca^{2} \ , \ee
so $\nabla_{a}\ca$ is timelike for $A^{2} < 1$.

For the class of $G_{2}$ cosmologies in which the spacetime
gradient $\nabla_{a}\ca$ is {\em timelike\/},  we can choose the
 gauge condition
\be \lb{tlareagc} A = 0 \ , \ee
which would be achieved by choosing $\p_{0}$ to be parallel to
$\nabla_{a}\ca$. It follows from equation. (\r{dad1}) that $\ptl_{x}\ca
= 0$, and we obtain $\ca = \ell_{0}^{2}\,\e^{2\cn_{0}t} $, which
is function of $t$ only (the so-called area time coordinate). This is 
refered to as {\em  the timelike area gauge} \cite {elst}.

There are other gauge choices, such as the fluid comoving gauge or the 
synchronons gauge, but these are less convenient for numerical analysis.

\subsection{Governing equations in timelike area gauge}
\lb{subsec:gst}

Let us explicitly give the evolution system in the timelike area
gauge using equations (\r{sgfix}), (\r{areagc}), (\r{areagn}) and 
(\r{tlareagc}). In the timelike area gauge we can use equation (\ref{sgfix}) to  eliminate the 
evolution equation (\ref{dlnpdot}), and equation (\ref{dladot}) becomes 
trivial. The relevant equations are:

\noindent {\em Evolution system\/}:

\bea \lb{Edlbetadot} \ptl_{t}B
& = & -(q+1)B \\
\lb{Edle11dot} \ptl_{t}E_{1}{}^{1}
& = & (q+3\Sigp)\,E_{1}{}^{1} \\
\lb{Edlsigpdot}
\ptl_{t}\Sigp
& = & (q+3\Sigp - 2)\,\Sigp-2\,(\Nm^2 + \Nc ^2)-\sfrac{1}{3}\,E_1{}^1\,\ptl_{x}r - \sfrac{3}{2}\,\Pi_{\text{total}}\\
\lb{Edlsigmdot} \ptl_{t}\Sigm + E_{1}{}^{1}\,\ptl_{x}\Nc & = &
(q+3\Sigp-2)\,\Sigm + 2\sqrt{3}\,\Sigc^{2}
- 2\sqrt{3}\,\Nm^{2} \\
\lb{Edlncdot} \ptl_{t}\Nc + E_{1}{}^{1}\,\ptl_{x}\Sigm
& = & (q+3\Sigp)\,\Nc \\
\lb{Edlsigcdot} \ptl_{t}\Sigc - E_{1}{}^{1}\,\ptl_{x}\Nm
& = & (q+3\Sigp-2-2\sqrt{3}\Sigm)\,\Sigc - 2\sqrt{3}\,\Nc\,\Nm \\
\lb{Edlnmdot} \ptl_{t}\Nm - E_{1}{}^{1}\,\ptl_{x}\Sigc & = &
(q+3\Sigp+2\sqrt{3}\Sigm)\,\Nm + 2\sqrt{3}\,\Sigc\,\Nc \eea
\bea \lb{Edlmudot} \frac{f_{1}}{\Om}\,(\ptl_{t} +
\frac{\gam}{G_{+}}\,v\,E_{1}{}^{1}\,\ptl_{x})\,\Om +
f_{2}\,E_{1}{}^{1}\,\ptl_{x}v & = &
2\,\frac{\gam}{G_{+}}\,f_{1}\,[\ \frac{G_{+}}{\gam}\,(q+1)
- \,\frac{1}{2}\,(1-3\Sigp)\,(1+v^2) - \ 1 \,] \\
\lb{Edlvdot} \frac{f_{2}}{f_{1}}\,\Om\,(\ptl_{t} -
\frac{f_{3}}{G_{+}G_{-}}\,v\,E_{1}{}^{1}\,\ptl_{x})\,v +
\,f_{2}\,E_{1}{}^{1}\,\ptl_{x}\Om & = &
-\,\frac{f_{2}}{f_{1}G_{-}}\,\Om\,(1-v^{2})\,
[\ \frac{(2-\gam)}{\gam}\,G_+\,r \nonumber\\
& & \hspace{15mm} + \,(2-\gam)\,(1-3\Sigp)\,v - 2\,(\gam-1)\,v  \,]  
\\
\lb{Edlqxdot}
        \ptl_t \Omega_u + \EEE \ptl_x Q_u
        &=& \left[2(q-1)+12\frac{1+v^2}{3+v^2}\Sigp\right]\Omega_u +vX
\\ 
\lb{Edlqdot}
        \ptl_t Q_u + \EEE \ptl_x (P_u + \Pi_u)
        &=& 2(q+3\Sigp-1)Q_u -\frac{2(1-v^2)}{3+v^2} r \Omega_u +X
\eea
where
\bea
        \ptl_x (P_u + \Pi_u) &=&
        \frac{1+3v^2}{3+v^2} \, \ptl_x \Omega_u
                + \frac{16 v}{(3+v^2)^2}\Omega_u \, \ptl_x v
\\
        X &\equiv& \frac{\gamma(1-v^2) B^2 \Omega^2}{6 {G_+}^2 G_-}
        \Big[ \gamma v (1-v^2)(1-3\Sigp) +2\gamma v +  2(1-v^2) r \Big]
\nonumber\\
        &&\quad - \frac{\gamma (1-v^2)^2}{6{G_+}^2G_-} B^2 \Omega
        \,\EEE \ptl_x \Omega
        + \frac{\gamma^2v(1-v^2)(3+(\gamma-1)v^2)}{6{G_+}^3G_-}
                B^2 \Omega^2 \,\EEE \ptl_x v
\eea

{\em Constraint equations\/}:
\bea
\lb{Efried} 0 & =&  2\Sigp -
1 + \Sigm^2 + \Sigc^2  +
\Nc^{2} + \Nm^{2}+\Om + \frac{(1-v^2)}{12\,G_+{}^2}\,\left(1+(2\gam-1)\,v^2\right)\Om^2 B^2 +\, \Omega_u \\
\lb{constraint2} 0 & = & r + 3\,(\Nc\,\Sigm-\Nm\,\Sigc) +\sfrac{3}{2} \left(\frac{\gam\,v}{G_+}\Om +  \frac{\gam\,v\,(1-v^2)}{6\,G_+{}^2}\Om^2\,B^2 + Q_u  \right) 
\eea
\noindent {\em Defining equations for $q$ and $r$\/}:
\bea \lb{Ehdecel} q & := & \sfrac{1}{2}
+ \sfrac{3}{2}\,(\Sigm^{2}+\Nc^{2}+\Sigc^{2}+\Nm^{2}) \nonumber\\
& & + \sfrac{3}{2}\,\left(  \frac{2\gam-1+v^2}{12\,G_+{}^2}\,(1-v^2)\,\Om^2\,B^2 +\,\frac{1+3\,v^2}{3+v^2}\Om_u + \frac{\gam-1+v^2}{G_+}\Om  \right)  \\
\lb{Eheq3sigp}
q+3\,\Sigp &=& 2 +\sfrac{3}{2}(-\Om_{\text{total}} + P_{\text{total}} +\Pi_{\text{total}}) \nonumber\\
 &=& 2+ 3\,(1-v^2)\,\left(\frac{(\gamma-1)(1-v^2)\Om^2B^2}{12\,G_+{}^2}-\frac{\Om_u}{3+v^2} -\frac{(2-\gamma)\Om}{2\,G_+}\right)\\
\lb{Ecodac1} r & := & -\frac{E_1{}^1\ptl_{x} B}{B}\ .
\eea
There is numerical (section \ref{sectionnumerical}) and dynamical (section \ref{sectionanisotropic}) evidence for the existance of a number of monotonic functions close to the initial singularity. In particular, from equations  (\r{Edlbetadot}) and (\ref{Ecodac1}) (using $(3 \gamma -1) \ge 0$), close to an isotropic singularity  $\beta$ is itself monotonic.

\section{Numerical Results}\label{sectionnumerical}

We have written the governing equations as a system of evolution 
equations subject to the constraint equations (\ref{Efried}) and 
(\ref{constraint2}). We can use (\ref{Efried}) to obtain $\Om_u$ and
(\ref{constraint2}) to solve for $Q_u$ and thus treat the governing system  
as a system of evolution equations without constraints (i.e., we don't need to use the evolution equations for $\Om_u$ and $Q_u$ in the numerics).  In the numerical analysis we use the standard CLAWPACK package for 
PDEs with one space variable (see \cite{finite} for background).
In the numerical calculations we prescribe periodic boundary conditions 
(we also implement Roe-averaging for the Riemann solver and choose 
Godunov splitting for source term splitting).

From the numerical analysis we find that the area expansion rate increases 
without bound ($\beta \rightarrow \infty$)  and the normalized frame 
variable \cite{elst} vanishes ($E_1{}^1 \rightarrow 0$) as $t \rightarrow 
-\infty$. Since $\beta \rightarrow \infty$ (and hence the Hubble rate 
diverges), there always exists an initial singularity. In addition, we 
find that $\{\Om, \Sigm,\Nc, \Sigc, \Nm, r\}\rightarrow 0 $ as 
$t\rightarrow -\infty$ for all $\gam>1$. In the case $\gam >4/3$, the 
numerics indicate that $\{v, \Om_u, Q_u, \Sigp\} \rightarrow 0$ (and $\Om_b 
\rightarrow 1$) for {\em all} initial conditions. In the case of 
radiation ($\gam=4/3$), the models still isotropize as $t\rightarrow 
-\infty$, albeit slowly. We shall investigate this degenerate case in 
more detail. For $\gam<4/3$, $\{v, \Om_u, Q_u, \Sigp\}$ tend to constant 
but non-zero values as $t\rightarrow -\infty$. It is interesting to note 
that $v^2 \not \rightarrow 1$; i.e., the tilt does not tend to an extreme 
value.

\begin{figure}[h!]
\centerline{\epsfig{figure=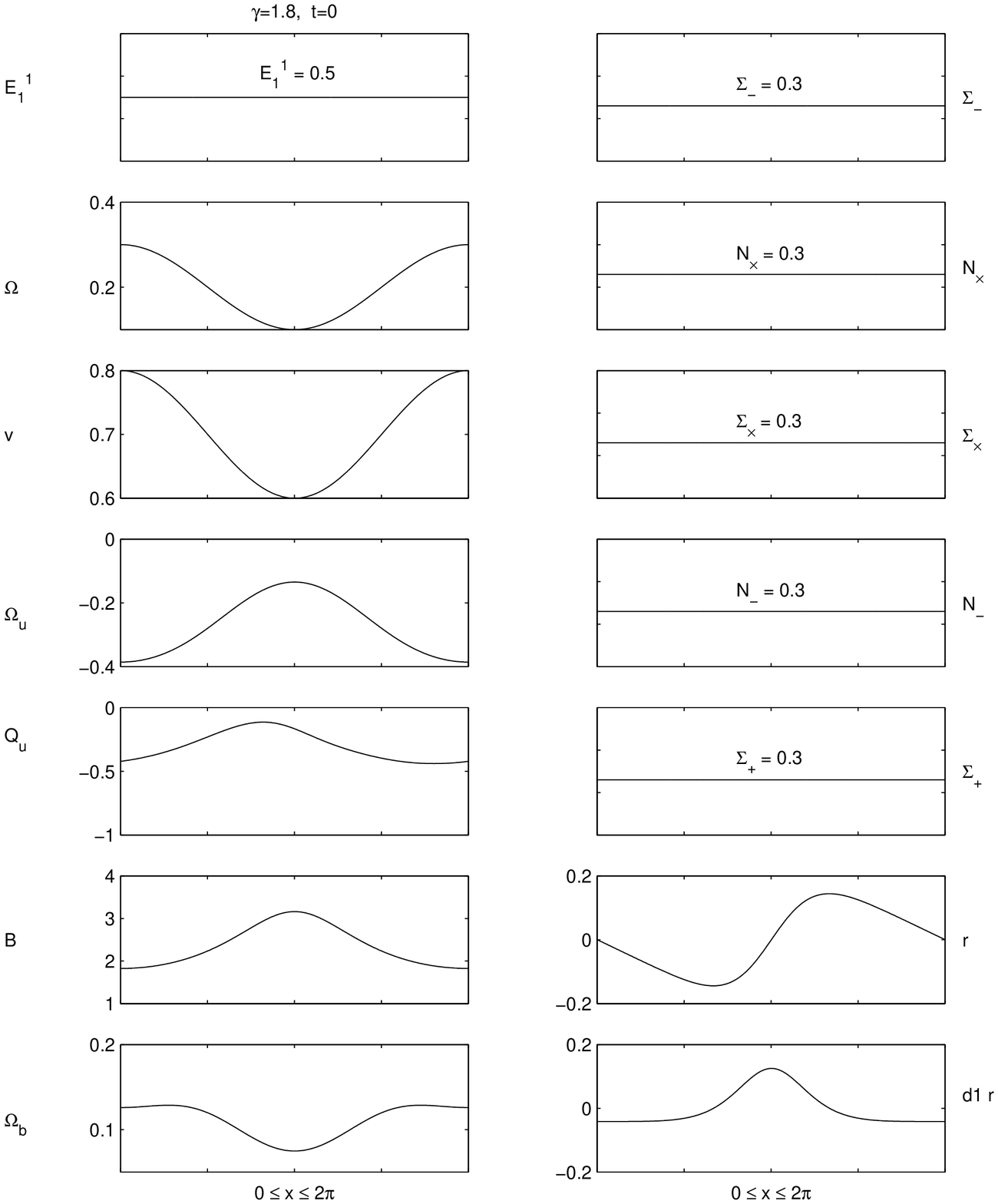, height=22cm,width=18cm}}
\caption{Isotropic singularity to the past for $\gam > 4/3$: $\gamma=1.8$, 
$t=0$}\label{G171}
\end{figure}

\begin{figure}[h!]
\centerline{\epsfig{figure=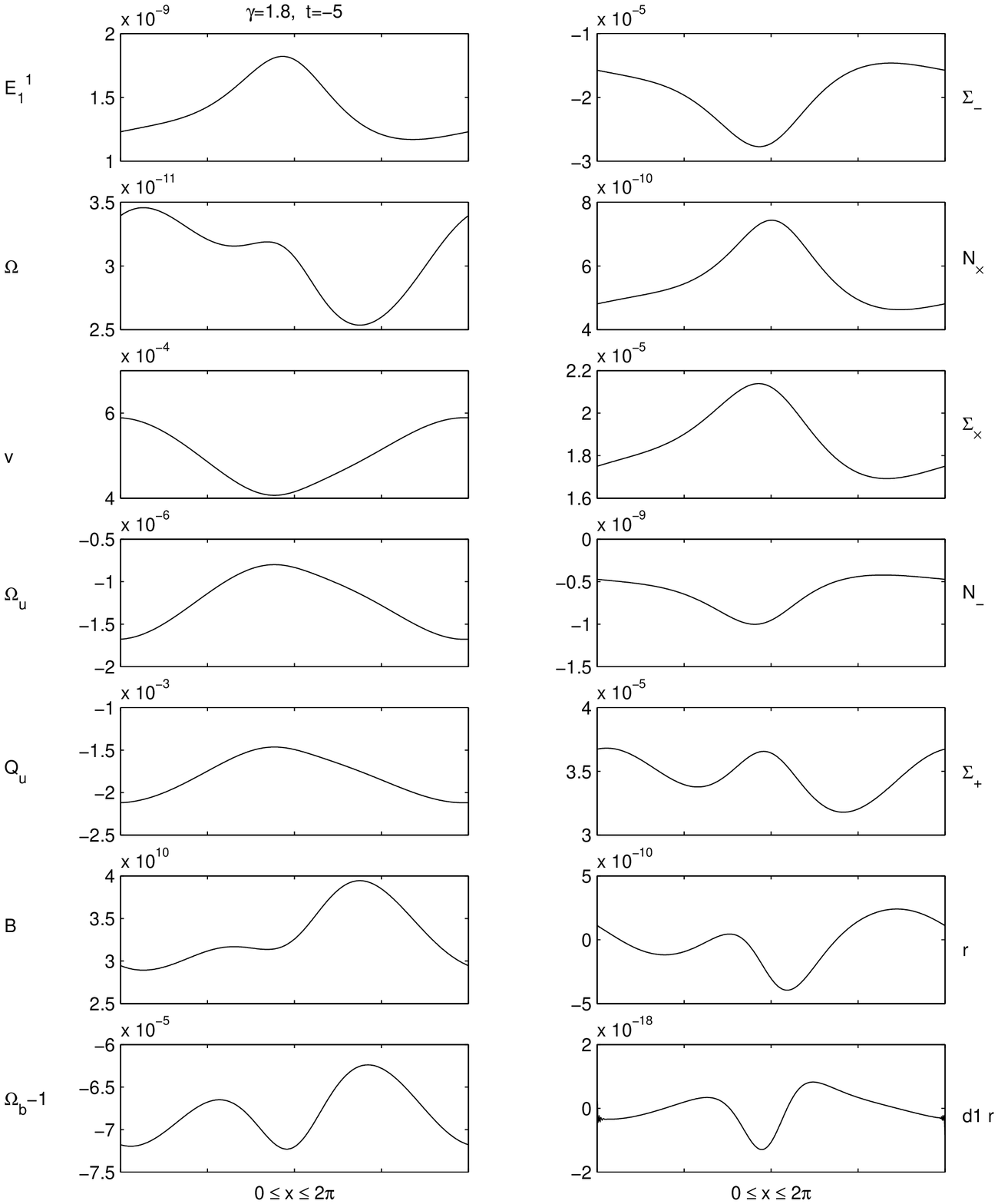, height=22cm,width=18cm}}
\caption{Isotropic singularity to the past for $\gam > 4/3$: $\gam=1.8$, $t=-5$}\label{G173}
\end{figure}

\begin{figure}[h!]
\centerline{\epsfig{figure=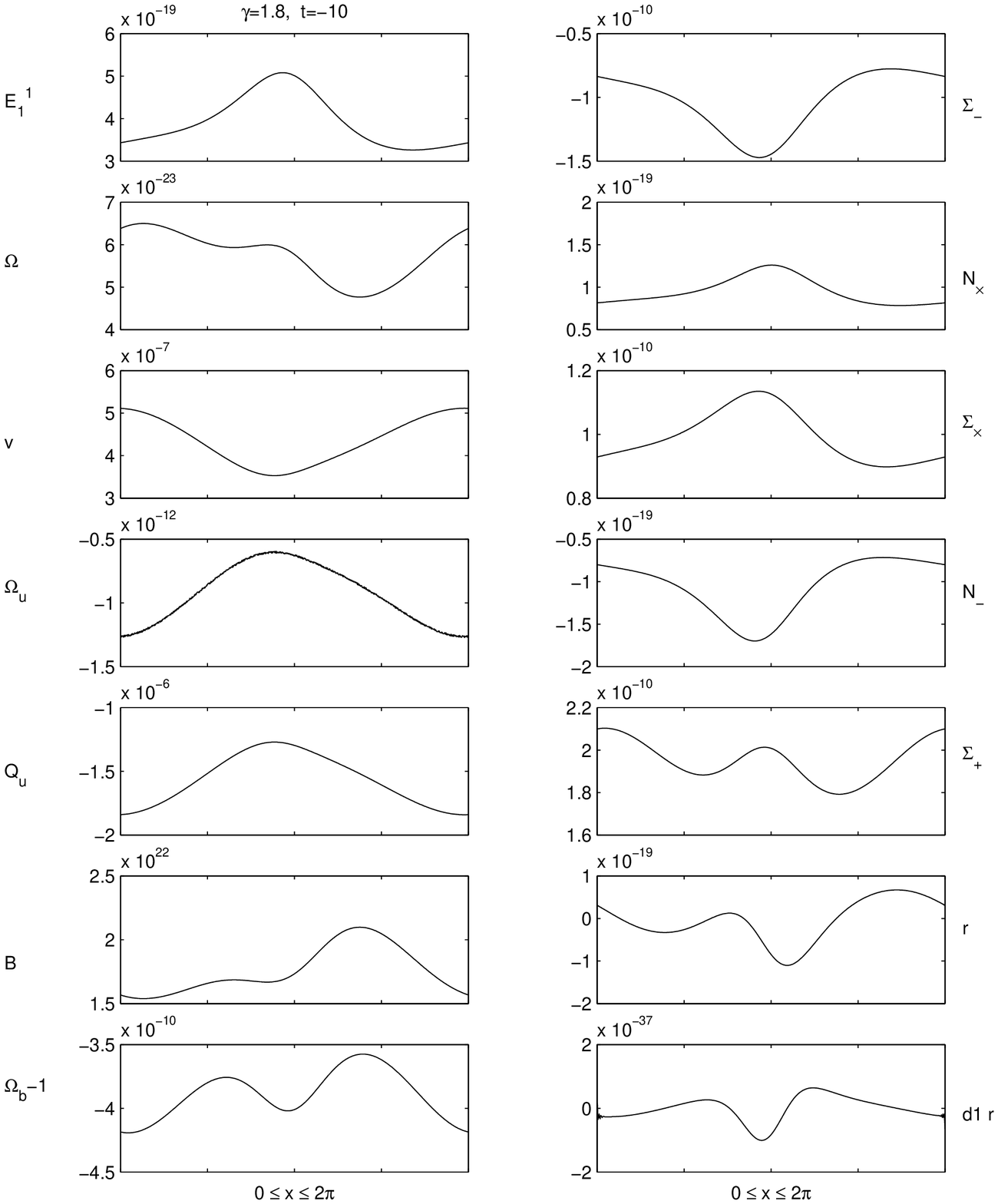, height=22cm,width=18cm}}
\caption{Isotropic  singularity to the past for $\gam > 4/3$:  $\gam=1.8$,  $t=-10$}\label{G175}
\end{figure}

\begin{figure}[h!]
\centerline{\epsfig{figure=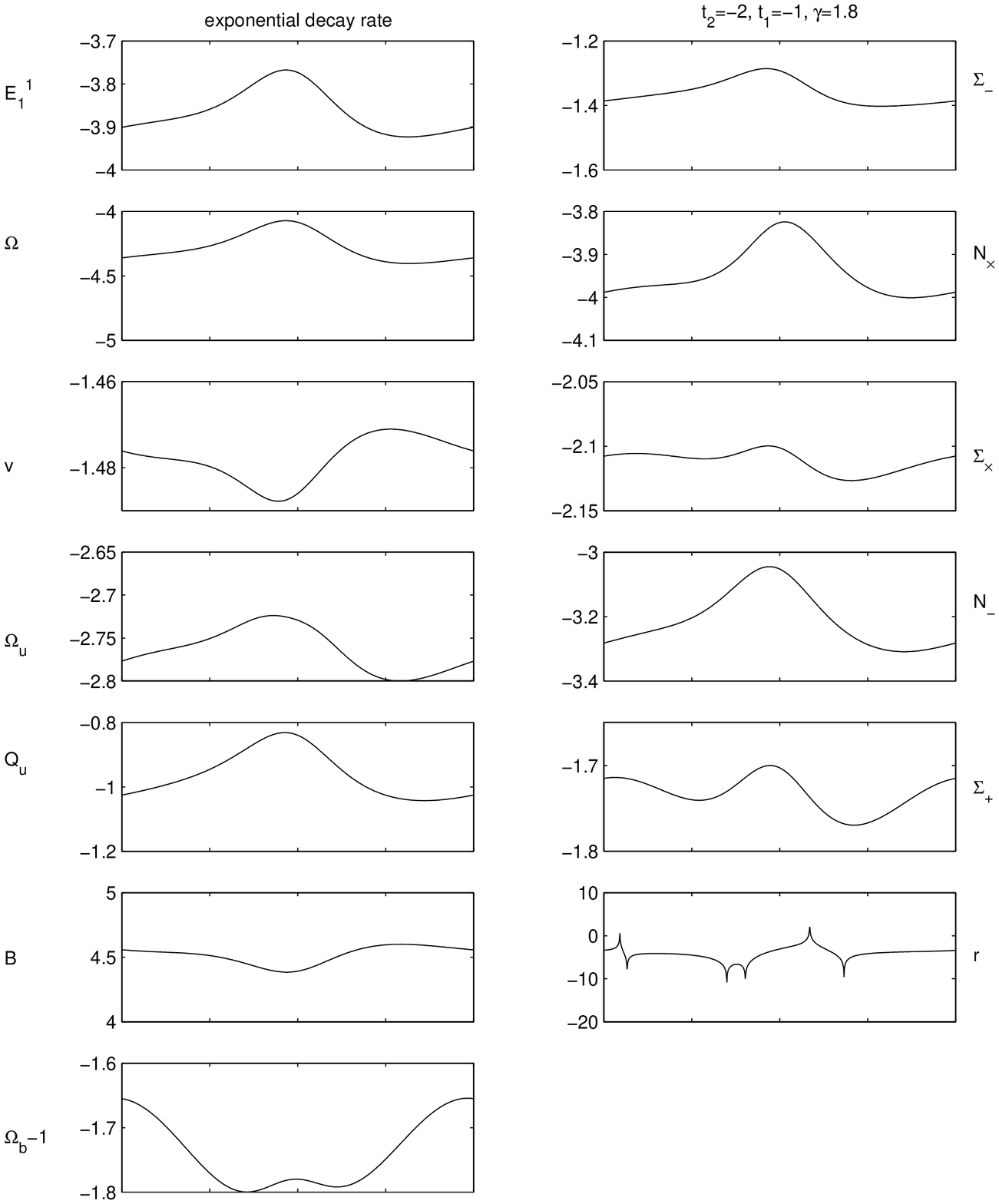,height=22cm,width=18cm}}
\caption{Exponential decay rates: $\gam=1.8$, $t_1=-1$ and  
$t_2=-2$}\label{dG170}
\end{figure}

\begin{figure}[h!]
\centerline{\epsfig{figure=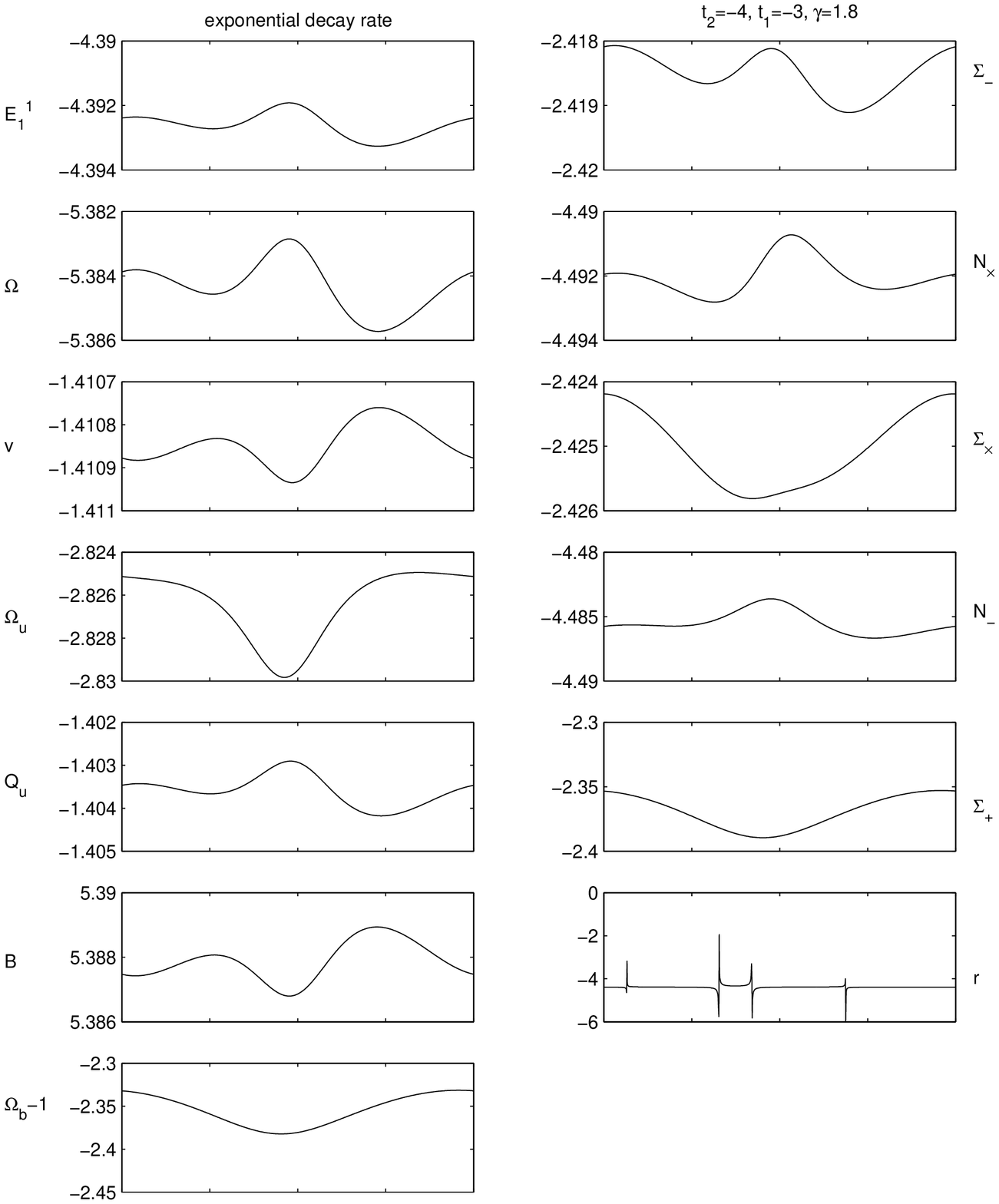,height=22cm,width=18cm}}
\caption{Exponential decay rates: $\gam=1.8$,  $t_1=-3$ and  
$t_2=-4$}\label{dG1704}
\end{figure}

The numerical results support the fact that {\em  all} cosmological models
 have an initial singularity and that for the range of values of
the equation of state parameter {\em  $\gamma > 4/3$} the singularity is 
isotropic. Indeed, the singularity is isotropic for all initial 
conditions (and not just for models close to ${\cal F}_b$) indicating 
that for $\gam >4/3$ this is the global behaviour.  We illustrate the 
numerical results for $\gamma=1.8$ by presenting some graphs of a typical  
run (see FIGs.\ref{G171}-\ref{G175}; in the TABLEs $d1 r$ denotes $\ptl_x\, r$) which show snapshots of the initial 
conditions and at earlier times, indicating isotropization. The numerical 
results support the exponential decay (to the past) of the anisotropies of 
section \ref{sectiondecayrate} and  \cite{chaos} (see 
FIGs.\ref{dG170}-\ref{dG1704}). Indeed, we found no evidence that models 
do not isotropize to the past for  $\gamma > 4/3$ \cite{bruni}.

We  noted in \cite{chaos},  that an inhomogeneous energy density with a non-zero
$\D_\mu{\rho}$ acts as a source for ${\cal Q}_{\mu}$, and we must check
that the evolution of ${\cal Q}_{\mu}$ is consistent with the approximations used
and consequently we have a 
self-consistent solution  close to ${\cal F}_b$
(again we note that physically  ${\cal P}_{\mu\nu}$ corresponds to graviational waves and
will likely not affect the the dynamics close to the singularity).
Suffice it to say that the numerical results discussed here indicate a self-consistent
solution and serve to justify the dynamical arguments in \cite{chaos}.

\begin{figure}[h!]
\centerline{\epsfig{figure=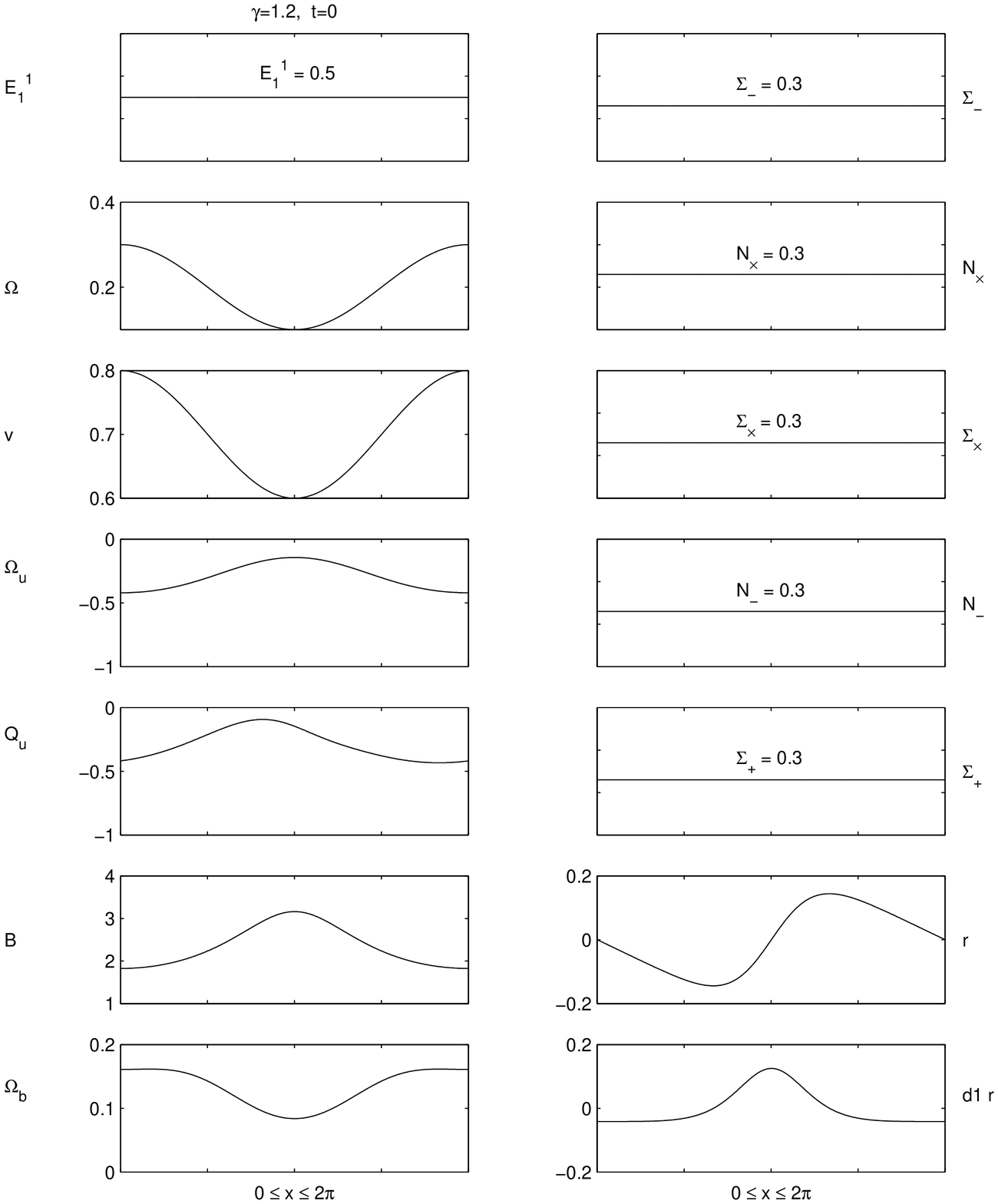, height=22cm,width=18cm}}
\caption{Initial singularity for $\gam < 4/3$: $\gam=1.2$,  
$t=0$}\label{G121}
\end{figure}

\begin{figure}[h!]
\centerline{\epsfig{figure=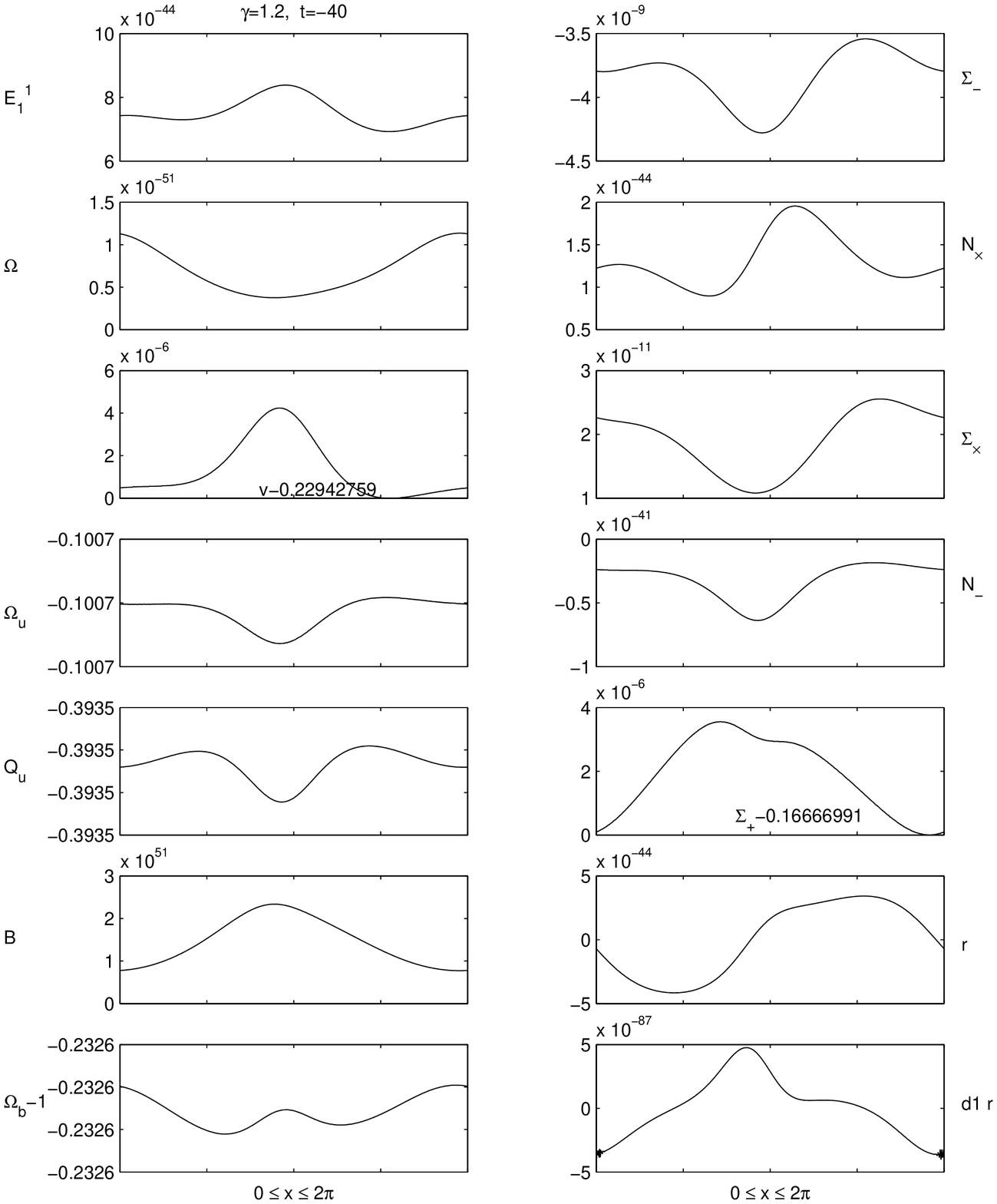,height=22cm,width=18cm}}
\caption{Initial singularity for $\gam < 4/3$: $\gam=1.2$, $t=-40$}\label{G122}
\end{figure}

\begin{figure}[h!]
\centerline{\epsfig{figure=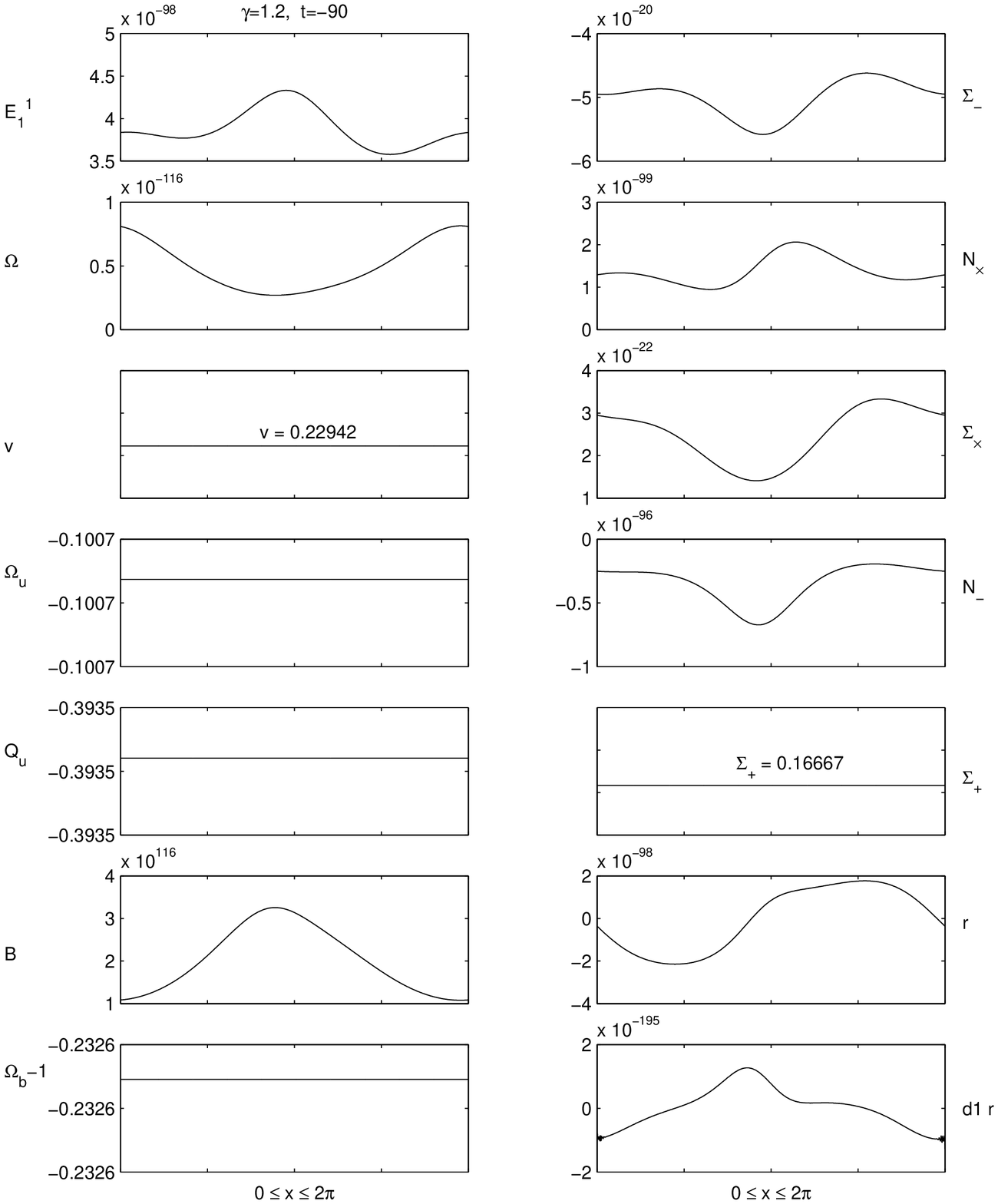,height=22cm,width=18cm}}
\caption{Initial singularity for $\gam < 4/3$: $\gam=1.2$,  $t=-90$}\label{G123}
\end{figure}

In the case $\gam < 4/3$ we find numerically that $\beta^{-1} \rightarrow 0$, so that there is always an initial singularity as $t \rightarrow -\infty$. In addition, we find that $E_1{}^1$, $r$, $\Nm$, $\Nc$, $\Sigm$ $\Sigc$, $\Om$ all vanish as the initial singularity is approached (see FIGs. \ref{G121}-\ref{G123}). However, the initial singularity is not, in general, isotropic. We shall discuss this, and the degenerate case $\gam =4/3$, further in sections \ref{sectiondecayrate} and \ref{sectionanisotropic}.

\section{derivation of asymptotic dynamics}\label{sectiondecayrate}

From the numerical analysis we have the following conditions when $\gam > 4/3$:
\bea
\lb{cond1}
C_1: & &\lim_{t\rightarrow -\infty} (\EEE, B^{-1}, r,
        \Nm, \Nc, \Sigm, \Sigc, \Sigp, \Om, v, \Om_u, Q_u, \Om_b-1 
        )
        = \mathbf{0}, \nonumber
\\
\lb{cond2}
C_2: & &\ptl_x (\EEE, B^{-1}, r,
        \Nm, \Nc, \Sigm, \Sigc, \Sigp, \Om, v, \Om_u, Q_u, \Om_b-1
        )
         \,\,\, \text{are bounded as} \,\,\,  t \rightarrow -\infty.\nonumber
\\
\lb{cond3}
C_3: & & V= {\cal O} (f(t)) \,\,\,\text{implies}\,\,\, \ptl_x V = 
{\cal O} (f (t)) \,\,\, \text{(asymptotic expansions in time}\nonumber
\\
        & &\text{can be differentiated with respect to the
        spatial coordinates)}.\nonumber
\eea
In particular, since $E_1{}^1 \rightarrow 0$ as $t \rightarrow -\infty$, we can follow the analysis in \cite{lim0306118} and use equations (\ref{Edlbetadot})-(\ref{Ecodac1}) to obtain the following asymptotic decay rates.\\

\noindent \textbf{Stage 1}: First, $C_1$ and $C_2$ imply that $q \rightarrow 3\gam-1$.
Using $C_1$, $C_2$ and the evolution equations, we obtain from 
Proposition 1 in \cite{lim0306118} in succession,
\bea
	\EEE &=& {\cal O}(\e^{[(3\gam-1)-\eps]t})
\\
\lb{Brate}
	B^{-1} &=& {\cal O}(\e^{[3\gam-\eps]t})
\\
	\Nc &=& {\cal O}(\e^{[(3\gam-1)-\eps]t})
\\
	\Nm &=& {\cal O}(\e^{[(3\gam-1)-\eps]t})
\\
	\Sigc &=& {\cal O}(\e^{[3(\gam-1)-\eps]t})
\\
	\Sigm &=& {\cal O}(\e^{[3(\gam-1)-\eps]t})
\eea
for any $\epsilon >0$.
It then follows from $C_1$, $C_2$, equation (\ref{Brate}) and the  algebraic 
expression for $\Om_b$ that
\be
	\Om = {\cal O}(\e^{[3\gam-\eps]t})\ .
\ee

\noindent \textbf{Stage 2}: Using $C_1$, $C_2$  and $C_3$, we obtain from equation (\ref{betar}) that
\be
	r = {\cal O}(\e^{[(3\gam-1)-\eps]t})\ .
\ee
Using $C_1$, $C_2$, $C_3$ and the evolution equations, we obtain from
Proposition 1 in succession,
\bea
        v &=&  {\cal O}(\e^{[(3\gam-4)-\eps]t})
\\
        \Sigp &=&  {\cal O}(\e^{[3(\gam-1)-\eps]t}
                                +\e^{[2(3\gam-4)-\eps]t})
\\
        Q_u &=&  {\cal O}(\e^{[(3\gam-4)-\eps]t})
\\
        \Om_u &=&  {\cal O}(\e^{[2(3\gam-4)-\eps]t})\ .
\eea

\noindent \textbf{Stage 3}:
First,
\be
	q = 3\gam-1 + g
\ee
where $g={\cal O}(\e^{[3(\gam-1)-\eps]t}
                        + \e^{[2(3\gam-4)-\eps]t})$, the dominant 
terms being $\Sigp$ and $v^2$.
As in \cite{lim0306118}, we use  $C_1$, $C_2$, $C_3$ and Proposition 4 on the 
evolution equations to obtain
\bea
	\EEE &=& \e^{(3\gam-1)t} [ \hat{E}_1{}^1 +{\cal O}(g) ]
\\
	B^{-1} &=& \e^{3\gam t} [ \hat{B}^{-1} + {\cal O}(g) ]
\\
	\Nc &=& \e^{(3\gam-1)t} [ \hat{N}_\times + {\cal O}(g) ]
\\
        \Nm &=& \e^{(3\gam-1)t} [ \hat{N}_- + {\cal O}(g) ]
\\
	\Sigc &=& \e^{3(\gam-1)t} [ \hat{\Sig}_\times + {\cal O}(g) ]
\\
        \Sigm &=& \e^{3(\gam-1)t} [ \hat{\Sig}_- + {\cal O}(g) ]
\\
	\Om &=& \e^{3\gam t} [ \sqrt{12} \hat{B}^{-1}  + {\cal O}(g) ]\,,
\eea
where the hat variables are functions of $x$ only.
Then (\ref{betar}) gives
\be
	r = \e^{(3\gam-1)t} [ - \hat{E}_1{}^1 
		\frac{\ptl_x \hat{B}}{\hat{B}} + {\cal O}(g) ]    \ .
\ee
The evolution equation for $v$ gives
\be
	v = \hat{v} \e^{(3\gam-4)t} + h\ ,
\ee
where $ h = {\cal O}(\e^{(3\gam-1)t}
                + \e^{[3(3\gam-4)-\eps]t})$, the dominant terms being 
$r$ and $v^3$.
The evolution equation for $\Sigp$ gives
\be
	\Sigp =\left\{ \begin{array}{ll}
		\hat{\Sig}_+ \e^{3(\gam-1)t}
		-\frac{2\gam}{3\gam-5} \hat{v}^2 \e^{2(3\gam-4)t}  
		+ {\cal O}(vh)
		& \text{$\frac{4}{3} < \gamma < 2, \gamma \neq \frac{5}{3}$}
		\\
		\frac{-10}{3} \hat{v}^2 t \e^{2t}
		+ \hat{\Sig}_+ \e^{2t} + {\cal O}(vh)
		& \text{ $\gam = \frac{5}{3}$.}
		\end{array} \right.
\ee
The evolution equation of $Q_u$ gives
\begin{equation}
	Q_u = -2\gam \hat{v} \e^{(3\gam-4)t} 
		+  \hat{Q}_u \e^{2(3\gam-2)t} + {\cal O}(h)\ .
\end{equation}
The evolution equation of $\Om_u$ gives
\begin{equation}
        \Om_u = -\frac{3}{2} \gam^2 \hat{v}^2 \e^{2(3\gam-4)t}
                + \hat{\Om}_u \e^{2(3\gam-2)t}
                +{\cal O}(vh)\ .   
\end{equation}
We then eliminate the epsilons from $g$ and $h$:
\be
	g = {\cal O}(\Sigp) = \left\{ \begin{array}{ll}
		{\cal O}(\e^{3(\gam-1)t}+\e^{2(3\gam-4)t})
                & \text{$\frac{4}{3} < \gamma < 2, \gamma \neq \frac{5}{3}$}
		\\
		{\cal O}(t \e^{2 t})
		& \text{ $\gamma = \frac{5}{3}$}
                \end{array} \right.
\ee
and
\begin{equation}
	h = {\cal O}(r+v^3) = {\cal O}(\e^{(3\gam-1)t}
				+ \e^{3(3\gam-4)t})\ .
\end{equation}

\textbf{Summary}: we have isotropization as $t \rightarrow -\infty$ for $\gam>4/3$ 
with the asymptotic decay rates:
\bea
        \EEE &=& \e^{(3\gam-1)t} [ \hat{E}_1{}^1 + {\cal O}(g) ]
\\
        B^{-1} &=& \e^{3\gam t} [ \hat{B}^{-1} + {\cal O}(g) ]
\\
       r &=& \e^{(3\gam-1)t} [ - \hat{E}_1{}^1
                \frac{\ptl_x \hat{B}}{\hat{B}} + {\cal O}(g) ]    \ .
\\
        \Nc &=& \e^{(3\gam-1)t} [ \hat{N}_\times + {\cal O}(g) ]
\\
        \Nm &=& \e^{(3\gam-1)t} [ \hat{N}_- + {\cal O}(g) ]
\\
\label{Sigcasy}
        \Sigc &=& \e^{3(\gam-1)t} [ \hat{\Sig}_\times + {\cal O}(g) ]
\\
\label{Sigmasy}
        \Sigm &=& \e^{3(\gam-1)t} [ \hat{\Sig}_- + {\cal O}(g) ]
\\
\label{Sigpasy}
        \Sigp &=& \left \{ \begin{array}{ll}
                \hat{\Sig}_+ \e^{3(\gam-1)t}
                -\frac{2\gam}{3\gam-5} \hat{v}^2 \e^{2(3\gam-4)t}
                + {\cal O}(vh)
                & \text{$\frac{4}{3} < \gamma < 2, \gamma\neq \frac{5}{3}$}
                \\
                \frac{-10}{3} \hat{v}^2 t \e^{2t}
                + \hat{\Sig}_+ \e^{2t} + {\cal O}(vh)
                & \text{ $\gamma = \frac{5}{3}$}
                \end{array} \right.
\\
\label{Omasy}
        \Om &=& \e^{3\gam t} [\sqrt{12} \hat{B}^{-1}  + {\cal O}(g) ]
\\
\label{vasy}
        v &=& \hat{v} \e^{(3\gam-4)t} + {\cal O}(h)
\\
        Q_u &=& -2\gam \hat{v} \e^{(3\gam-4)t}
                +  \hat{Q}_u \e^{2(3\gam-2)t} + {\cal O}(h)
\\
        \Om_u &=& -\frac{3}{2} \gam^2 \hat{v}^2 \e^{2(3\gam-4)t}
                + \hat{\Om}_u \e^{2(3\gam-2)t}
                +{\cal O}(vh)\ ,
\eea
where
\bea
        g &= & {\cal O}(\Sigp) = \left \{ \begin{array}{ll}
                 {\cal O} (\e^{3(\gam-1)t}+\e^{2(3\gam-4)t})
                & \text{$\frac{4}{3} < \gamma < 2, \gamma \neq \frac{5}{3}$}
                \\
                {\cal O}(t \e^{2t})
                & \text{ $\gamma = \frac{5}{3}$}
                \end{array} \right.
\\
        h &=& {\cal O}(r+v^3) = {\cal O}(\e^{(3\gam-1)t}
                                + \e^{3(3\gam-4)t})\ .
\eea

The asymptotic decay rates can be calculated numerically by plotting the 
log ratio of variables at different times. The case $\gam = 1.8$ is 
presented in FIGs \ref{dG170}-\ref{dG1704}. Note that the numerical 
calculations are consistent with the decay rates derived above.\\

Next let us discuss  the asymptotic dynamics when $\gam=4/3$. 
Suppose the conditions $C_1$, $C_2$ and $C_3$ again hold. Then, since $E_1{}^1 \rightarrow 0$ as $t \rightarrow -\infty$, we can again follow the analysis in \cite{lim0306118} and use equations (\ref{Edlbetadot})-(\ref{Ecodac1}) to obtain the asymptotic decay rates. Again, $C_1$ and $C_2$ imply that $q \rightarrow 3$, whence we obtain from Proposition 1 in succession,
\bea
        \EEE &=& {\cal O}(\e^{(3-\eps)t})
\\
\label{Bbrate}
        B^{-1} &=& {\cal O}(\e^{(4-\eps)t})
\\
        \Nc &=& {\cal O}(\e^{(3-\eps)t})
\\
        \Nm &=& {\cal O}(\e^{(3-\eps)t})
\\
        \Sigc &=& {\cal O}(\e^{(1-\eps)t})
\\
        \Sigm &=& {\cal O}(\e^{(1-\eps)t})\ .
\eea
It then follows from $C_1$, $C_2$, $C_3$,  equations (\ref{betar}) and 
(\ref{Bbrate}) that
\bea
        \Om &=& {\cal O}(\e^{(4-\eps)t}) 
\\
        r &=& {\cal O}(\e^{(3-\eps)t})\ .
\eea
Unfortunately, $v$ does not decay exponentially and we cannot continue 
with the integration method of~\cite{lim0306118}.

However, we can still find the remaining decay rates heuristically. Equations 
(\ref{Edlvdot}) and (\ref{Edlsigpdot})  become asymptotically
\bea
	\ptl_t v &=& 2 \Sigp v + ...
\\
	\ptl_t \Sigp &=& \Sigp - \frac83 v^2 + ...\ ,
\eea
which give the approximate solutions
\bea
        v &=& \frac{1}{\sqrt{\frac{-32}{3}t + \frac{1}{v_0^2}}} + ...
\\
        \Sigp &=& \frac{8}{3} v^2 + \hat{\Sigma}_+ \e^{(1-\eps)t} + ...\ .
\eea
It follows that
\bea
        Q_u &=& - \frac{8}{3} v + \hat{Q}_u \e^{(4-\eps)t} + ...
\\
        \Omega_u &=& - \frac{8}{3}v^2 + \hat{\Omega}_u \e^{(4-\eps)t} +... 
\ .
\eea
These expressions are supported by the numerics.

\section{The invariant set $E_1{}^1=0$}\label{sectionanisotropic}

From numerical experiment we find that as $t \rightarrow -\infty$, 
$E_1{}^1 \rightarrow 0$ very rapidly,  $\Om,\,\Sigm,\, \Sigc, 
\Nc,\Nm,\,r\, \rightarrow 0$ and $B \rightarrow \infty$. We are 
now interested in the dynamics in the invariant set $E_1{}^1=0$ at 
early times. In order to find the early time behaviour of $\Omega_b$ (which 
is given in terms of $\Omega$, $B$ and $v$ by equation (\ref{defomb})) in 
the invariant set $\EEE=0$,  we need to obtain the evolution equation of 
$\Omega_b$ from the evolution equations of $\Omega$, $B$ and $v$ as follows:
\be
\partial_t \Omega_b = \partial_v \Om_b\,\partial_t v +  \partial_\Omega \Omega_b\,\partial_t \Omega +  \partial_B \Omega_b\,\partial_t B
\ee
where $\partial_t B$,  $\partial_t \Om$ and $\partial_t v$ are given 
separately by (\ref{Edlbetadot}), (\ref{Edlmudot}) and (\ref{Edlvdot}) 
with $E_1{}^1=0$. \\

Equations (\ref{Edlbetadot})-(\ref{Edlqdot}) imply that $E_1{}^1=\Sigm=\Sigc=\Nc=\Nm=\Om=0$ is an invariant subset of the invariant set $\EEE=0$. In this invariant subset
\be
 -\frac{3}{2}(Q_b+Q_u)=  r \equiv  -\frac{\EEE \ptl_x B}{B}=0
\ee
The dynamics in the invariant subset $E_1{}^1=\Sigm=\Sigc=\Nc=\Nm=\Om=0$ (and hence $r=Q_b+Q_u=0$) is given by
\bea
\lb{hedotOb}
\partial_t\Om_b &=& 2\left\{\frac{2+6\Om_b+6\Om_u-3\gam(1+\Om_b+\Om_u)}{(1+(2\,\gam-1)\,v^2)\,G_-\,G_+}\right\}\,\gam^2\,v^4\,\Om_b \nonumber\\
 && +2\,\Om_b\,\left\{(q+1) - 3(1-\Sigp)\,(1+\frac{3(\gam-1)(1-v^2)+\gam\,v^2}{3\,G_+}) +\Sigp\frac{2\gam v^2}{G_+}\right\}  \\
\lb{hedotv}
\partial_t v &=& -\frac{v(1-v^2)}{G_-}\,\left((2-\gam)(1-3\Sigp)-2(\gam-1)\right)  \\
\lb{hedotOu}
\partial_t\Om_u &=& 2(q+1)\,(\Om_b+\Om_u)-3(1-\Sigp)\left\{\Om_b+\Om_u+\frac{1}{3}\left(\frac{3(2\gam-1)(1-v^2)+2\gam\,v^2}{(1+(2\,\gam -1)\,v^2)}\right)\Om_b + \frac{1}{3}\Om_u\right\} \nonumber\\
 & & +3\Sigp\,\left\{\frac{4}{3}\frac{\gam\,v^2}{(1+(2\gam-1)v^2)}\Om_b +\frac{8\,v^2}{3\,(3+v^2)}\Om_u \right\} -2\left\{\frac{2+6\Om_b+6\Om_u-3\gam(1+\Om_b+\Om_u)}{(1+(2\,\gam-1)\,v^2)\,G_-\,G_+}\right\}\,\gam^2\,v^4\,\Om_b \nonumber\\
 && -2\,\Om_b\,\left\{(q+1) - 3(1-\Sigp)\,(1+\frac{3(\gam-1)(1-v^2)+\gam\,v^2}{3\,G_+}) +\Sigp\frac{2\gam v^2}{G_+}\right\} 
\eea
with
\bea
\lb{hesigp}
\Sigp &=& \frac{1}{2}\,(1-\Om_b-\Om_u)\\
\lb{heq}
q&=& \frac{1}{2} +\frac{1}{2}\,\left\{\frac{3(2\gam-1)(1-v^2)+2\gam\,v^2}{(1+(2\,\gam -1)\,v^2)}\Om_b + \Om_u +\frac{4\gam\,v^2}{(1+(2\gam-1)v^2)}\Om_b +\frac{8\,v^2}{(3+v^2)}\Om_u\right\}\\
\text{Note that} \ \ \ \ \ \ \ \ \ \ \ \ \ \ \ \ \ \ \ \ & &\\
\lb{hequ}
Q_u &=& -Q_b = -\frac{2\,\gam\, v}{1+(2\,\gam -1)\,v^2}\,\Om_b
\eea
Since $\Om_b \sim \Om^2\,B^2$ (see equation (\ref{defomb})), in this invariant set we have a closed system of ODE ((\ref{hedotOb})-(\ref{hedotOu})) in terms of the three variables $\Om_b$, $v$ and $\Om_u$.\\

To obtain the  equilibrium points, we let $\partial_t \Om_b=0$, 
$\partial_t v=0$ and $\partial_t \Om_u$ in equations 
(\ref{hedotOb})-(\ref{hedotOu}). The results are summarized in TABLE~\ref{eqpt}, where 

\begin{table}{\center
\caption{Equilibrium points and their stability} \label{eqpt}
\begin{tabular}{|l|l|l|}
\hline
equilibrium point& eigenvalues & stability  \\
\hline
$\Om_b=0,\, v=0,\, \Om_u=0$ & $-3\,(\gam-1),\, \frac{3}{2}\,\gam-1,\, 1$ & Saddle\\
\hline
$\Om_b=0,\, v=0,\, \Om_u=1$ & $-2\,(3\,\gam-2),\, 3\,\gam-4,\, -1$ & Saddle when $\gam> 4/3$. Sink when $\gam<4/3$\\
\hline
$\Om_b=1,\, v=0,\, \Om_u=0$ (${\cal F}_b$) & $3\,\gam-4,\, 3\,(\gam -1), \, 2\,(3\,\gam-2) $ & {\bf Source when $\gam> 4/3$. Saddle when $\gam<4/3$}\\
\hline
$\Om_b=0,\, v=\pm 1, \, \Om_u=0$ & $ 2,\, -\frac{3\,\gam -2}{2-\gam},\, -4\frac{\gam -1}{2-\gam}$& Saddle\\
\hline
$\Om_b=F_1(\gam),\, v=\pm F_2(\gam), \, \Om_u= F_3(\gam)$ (${\cal P}_\pm$) & $\lambda_1(\gam)>0,\, \lambda_2(\gam)>0,\, \lambda_3(\gam)>0$& {\bf $\gam<4/3$. Source}. \\
\hline
\end{tabular}}
\end{table}

\begin{table}{\center
\caption{Nonhyperbolic equilibrium point ${\cal F}_b={\cal P}_\pm$ when 
$\gam=4/3$} \label{degenerate}
\begin{tabular}{|l|l|}
\hline
\ \ \ \ \ The values of the variables& $\Om_b=1,\ \ \ v=0,\ \ \ \Om_u=0,\ \ \ Q_b=Q_u=\Sigp=0,\ \ \ q=3$\\
\hline
\ \ \ \ \ The eigenvalues and eigenvectors \ \ \ \ \ &$\lambda_1=1$, $v_1=(1,0,0)$.\ \ \ $\lambda_2=4$, $v_2=(-1,0,1)$.\ \ \  $\lambda_3=0$, $v_3=(0,1,0)$  \ \ \ \ \ \\
\hline
\end{tabular}}
\end{table}
\be
	F_1 \equiv 1 + \frac{(3\gam-4)(3\gam^3-21\gam^2+36\gam-16)}{
		3\gam(3\gam-2)(2-\gam)^2}\ ,\quad
	F_2 \equiv \sqrt{ \frac{(3\gam-4)(\gam-1)}{2\gam^2-7\gam+4}}
		\ ,\quad
	F_3 \equiv - \frac{(3\gam-4)(\gam-1)(9\gam^2-28\gam+16)}{
		3\gam(3\gam-2)(2-\gam)^2}\ .
\ee
For ${\cal P}_\pm$,  we also obtain the values of $q$, $\Sigp$, $Q_b$ and $Q_u$ from equations (\r{hesigp})-(\r{hequ}):
\bea
q &=& \frac{3\,\gam-2}{2-\gam}      \\
\Sigp &=&\frac{4-3\,\gam}{3(2-\gam)}\equiv F_4(\gam)  \\
Q_b &=& \pm \frac{2\,(3\,\gam^2-12\,\gam+8)\sqrt{(2\gam^2-7\gam+4)(3\,\gam -4)\,(\gam-1)}}{3\,\gam\,(3\,\gam -2)\,(2-\gam)^2}\equiv \pm F_5(\gam)\\
Q_u &=& \mp \frac{2\,(3\,\gam^2-12\,\gam+8)\sqrt{(2\gam^2-7\gam+4)(3\,\gam -4)\,(\gam-1)}}{3\,\gam\,(3\,\gam -2)\,(2-\gam)^2}=\mp F_5(\gam)
\eea
Let us comment on the equilibrium points ${\cal P}_\pm$. $F_2 (\gam)$ is a real number only when $\gam\leq 4/3$, whence $0<\Om_b\leq 1$, $0\geq \Om_u> -1$, $|v|<1$. So ${\cal P}_\pm$ exist only when  $\gam\leq 4/3$.   The expressions for $\lambda_i, \, i=1,2,3$ are very complicated; to determine stability we calculated  $\lambda_i, \, i=1,2,3$ for different values of $\gam < 4/3$ numerically and found that $Re(\lambda_i)>0$, for  $i=1,2,3$ when $1<\gam < 4/3$. We conclude that ${\cal P}_\pm$ are sources for the dynamics in the invariant subset given by $E_1{}^1=\Sigm=\Sigc=\Nc=\Nm=\Om=0 \,(=r =Q_b+Q_u)$ when $\gam<4/3$.\\

Therefore in the invariant subset given by $E_1{}^1=\Sigm=\Sigc=\Nc=\Nm=\Om=0$, ${\cal F}_b$ is the global source for $\gam>4/3$,  and ${\cal P}_\pm$ ($\pm$ depending on sign of $v$) are  anisotropic sources for $\gam<4/3$. There is a bifurcation at $\gam=4/3$. Let us next discuss this degenerate case. When $\gam=4/3$, ${\cal P}_\pm$ and ${\cal F}_b$ coincide. The values of the variables corresponding to this nonhyperbolic equilibrium point, the eigenvalues and  the corresponding eigenvectors are given in TABLE \ref{degenerate}. At this nonhyperbolic equilibrium point, there exists a 2-dimensional unstable manifold and a 1-dimensional center manifold. The 1-dimensional center manifold is tangent to $v_3=(0,1,0)$ at this point, therefore the center manifold has the form $\Om_b-1=a_0v^2+a_1v^3+a_2v^4+a_3v^5+{\cal O}(v^6)$ and $\Om_u=b_0v^2+b_1v^3+b_2v^4+b_3v^5+{\cal O}(v^6)$.  The stability is determined by the dynamics in the center manifold. We find that  the center manifold is g
 iven by $\Om_b-1=-\frac{8}{3}v^2-\frac{136}{3}v^4+{\cal O}(v^5)$ and $\Om_u=-\frac{8}{3}v^2+\frac{40}{9}v^4+{\cal O}(v^5)$ and that the dynamics in the center manifold is consequently determined by $\frac{dv}{dt}=\frac{16}{3}v^3+\frac{112}{3}v^5+{\cal O}(v^6)$. From the dynamics in the center manifold, we find that the nonhyperbolic equilibrium point is a source for the center manifold, therefore the nonhyperbolic equilibrium point is a source for the 3-dimensional invariant subset $E_1{}^1=\Sigm=\Sigc=\Nc=\Nm=\Om=0$. This is consistent with the decay rates calculated above and with numerical analysis (and phase portraits in the 3-dimensional invariant set). Therefore models isotropize `slowly' in the radiation case. In summary, when $\gam> 4/3$,  ${\cal F}_b$ is a global source; when $\gam=4/3$, ${\cal F}_b={\cal P}_\pm$ is still a global source;  when $\gam<4/3$, ${\cal F}_b$ becomes a saddle and a new pair of equilibrium points ${\cal P}_\pm$ appear as a pair of sources. 
 In addition to ${\cal F}_b$ and ${\cal P}_\pm$, there are also a number of saddle equilibrium points (see TABLE \ref{eqpt}).\\

We claim that when $\gam\geq 4/3$, ${\cal F}_b$ ($E_1{}^1=\Sigm=\Sigc=\Nc=\Nm=\Om=r =Q_b=Q_u=\Om_u=v=0,\, \Om_b=1$) is also a source for the full state space and  that when $\gam< 4/3$,  ${\cal P}_\pm$ ($E_1{}^1=\Sigm=\Sigc=\Nc=\Nm=\Om=r=\Sigp =0,\, \Om_b=F_1(\gam),\, v=\pm F_2(\gam),\, \Om_u=F_3(\gam),\, \Sigp=F_4(\gam),\, Q_b=\pm F_5(\gam),\, Q_u=\mp F_5(\gam)$) are  sources for the full state space. We give an argument for our claim  as follows. First, we observe that at ${\cal F}_b$ 
\[q+1=3\gam>0, \, \,\,\,\,\,
 q+3\Sigp-2=3(\gam-1)>0, \,\,\,\,\,\, T_1=\frac{3}{2}>0,\,\,\,\,\,\, T_2=3\gam-1>0.\] 
and  at ${\cal P}_\pm$ 
\[q+1=\frac{2\gam}{2-\gam}>0, \,\,\,\,\,\, q+3\Sigp-2=\frac{2(\gam-1)}{(2-\gam)}>0, \,\,\,\,\,\,  T_1=-\frac{(3\gam-2)\gam}{(2\gam^2-7\gam+4)}>0, \,\,\,\,\,\, T_2=\frac{2}{2-\gam}>0.\] 
where
\[T_1 \equiv \frac{(1+(\gam-1)v^2)}{\gam}(q+1)-\frac{1}{2}(1-3\Sigp)(1+v^2)-1,\ \ \ T_2 \equiv 2(q+3\Sigp-1)-\frac{3}{2}(-\Om_b-\Om_u+P_b+P_u+\Pi_b+\Pi_u).\]
So near the equilibrium points ${\cal F}_b$ and ${\cal P}_\pm$, $q+1>0$ and $q+3\Sigp-2>0$ and thus as $t\rightarrow -\infty$, we have $B\rightarrow +\infty$ and 
$E_1{}^1\rightarrow 0$.  Near ${\cal F}_b$ and ${\cal P}_\pm$, we can neglect the terms with `$E_1{}^1\partial_x$' in the equations (\ref{Edlbetadot})-(\ref{Edlqxdot}) and treat the PDE as a system of ODE (see section \ref{sectiondecayrate}). Then from the linearization of equations (\ref{Edlsigmdot})-(\ref{Edlmudot})  near  ${\cal F}_b$ and  ${\cal P}_\pm$ and using the fact that
$q+3\Sigp-2>0$, $T_1>0$ near  ${\cal F}_b$ and  ${\cal P}_\pm$, we find that as $t\rightarrow -\infty$, $\Sigm$, $\Sigc$, $\Nc$, $\Nm$ and $\Om$ (therefore $Q$) will decrease monotonically towards zero. We also observe that near ${\cal F}_b$ and ${\cal P}_\pm$
\bea
\partial_t (Q_b+Q_u)&=&2(q+3\Sigp-1)(Q_b+Q_u) \nonumber\\
 & & +\left(3(\Nm\Sigc-\Nc\Sigm)-\frac{3}{2}(Q+Q_b+Q_u)\right)(-\Om_b-\Om_u+ P_b+P_u+\Pi_b+\Pi_u)\\
&=& \left(2(q+3\Sigp-1)-\frac{3}{2}(-\Om_b-\Om_u+ P_b+P_u+\Pi_b+\Pi_u) \right)(Q_b+Q_u)  \nonumber\\
&& +\left(3(\Nm\Sigc-\Nc\Sigm)-\frac{3}{2}Q\right)(-\Om_b-\Om_u+ P_b+P_u+\Pi_b+\Pi_u)\,.
\eea
Linearizing this evolution equation for $Q_b+Q_u$ near ${\cal F}_b$ and ${\cal P}_\pm$ and using $T_2 >0$, we conclude that $Q_b+Q_u$, and thus $r$ ($=3(\Nm\Sigc-\Nc\Sigm)-\frac{3}{2}(Q+Q_b+Q_u)$), decrease towards zero as  $t\rightarrow -\infty$. \\

In summary, near  ${\cal F}_b$ and ${\cal P}_\pm$,  $E_1{}^1$, $\Sigm$, $\Sigc$, $\Nc$, $\Nm$,  $\Om$ and $r$ will decrease monotonically towards zero as $t\rightarrow -\infty$, so all orbits in the full state space evolve back towards the invariant subset given by  $E_1{}^1=\Sigm=\Sigc=\Nc=\Nm=\Om=0$. Because  ${\cal F}_b$ (when $\gam\geq 4/3$) and ${\cal P}_\pm$ (when $\gam<4/3$) are sources in this invariant subset, we expect that the orbits mentioned above will shadow orbits in the invariant subset and evolve back towards ${\cal F}_b$ (when $\gam\geq 4/3$) or ${\cal P}_\pm$ (when $\gam<4/3$). Therefore,  ${\cal F}_b$ is a source for the full state space when $\gam\geq 4/3$  and  ${\cal P}_\pm$  are sources for the full state space when $\gam<4/3$. The equilibrium points ${\cal P}_\pm$ correspond to new anisotropic brane world cosmologies. Note that the tilt is not extreme ($v^2\not=1$) at ${\cal P}_\pm $.\\

The numerical experiments have confirmed that ${\cal F}_b$ is a source in the full state space when $\gam \geq 4/3$ (see FIGs.\ref{G171}-\ref{G175}). The numerical simulation is consistent with the decay rates given in section 
\ref{sectiondecayrate} and with the analysis given above that ${\cal 
F}_b$ is a source for the full state space when $\gam=4/3$. Therefore 
models isotropize `slowly' in the radiation case. Next, we present numerical evidence that the ${\cal P}_\pm$ are sources when $\gam<4/3$. For example, when $\gam=1.2$, $\Om_b=0.767361111,\, v=\pm 0.2294157339,\, \Om_u= -0.1006944444$ (and  $\lambda_{1,2} =0.27236 \pm 0.4838767293 i,\, \lambda_3= 2.644927 $);  in this case $q=1.9999999$, $\Sigp=0.1666666$, $Q_b=\pm 0.3935117$ and $Q_u= \mp 0.3935117$. These results  for $\gam=1.2$ are consistent with the results given by numerical experiment (see FIGs.\ref{G121}-\ref{G123}). \\

\section{Discussion}
All models have an initial singularity as $t\rightarrow -\infty$. In 
addition, we find that $\{\Om, \Sigm,\Nc, \Sigc, \Nm, r\}\rightarrow 0 $ 
as $t\rightarrow -\infty$ for all $\gam>1$. In the case $\gam >4/3$, the 
dynamical and numerical analysis indicates that $\{v, \Om_u, Q_u, \Sigp\} 
\rightarrow 0$ (and $\Om_b \rightarrow 1$) for {\em all} initial 
conditions. In the case of radiation ($\gam=4/3$), the models still 
isotropize as $t\rightarrow -\infty$, albeit slowly. For $\gam<4/3$, 
$\{v, \Om_u, Q_u, \Sigp\}$ tend to constant but non-zero values as 
$t\rightarrow -\infty$.\\
 
\subsection{Tilt}
In the invariant set $v=0$, all models isotropize to the past (for $\gam>1$). Thus in the spatially homogeneous case with no tilt (with $\EEE=0$),  it follows that there exists an isotropic
singularity in all orthogonal Bianchi brane-world models in which
$\Omega_b$ dominates as the initial singularity is approached into the past (consistent with the results of \cite{COLEY}).
In particular, ${\cal F}_b$ is a local source and in general the initial singularity is isotropic.  The  linearized solution representing a general
solution in the neighbourhood of the initial singularity in a class of $G_2$ models was given in \cite{chaos}; it was found that ${\cal F}_b$ is a local source or 
past-attractor 
in  this family of spatially inhomogeneous cosmological models for $\gamma >1$. The exponential decay rates of the $G_2$ models (about ${\cal F}_b$) given in \cite{chaos} are consistent with those here in $v=0$ case (for all $\gam$).\\

Let us now assume that $v\not=0$ (the general case). For the $G_2$ models 
in the timelike area gauge we recover the orthogonally transitive
{\em tilting} Bianchi type VI$_0$  and VII$_0$ models in the spatially homogeneous limit 
(with one tilt variable). (A spatially homogeneous  cosmology is said to be {\it tilted}
if the fluid velocity vector is not orthogonal to the
group orbits, otherwise the model is said to be {\em
non-tilted}). There are no sources in the Bianchi type VI$_0$  and VII$_0$ models
with tilted perfect fluid; the past attractor is an infinite sequence
of orbits between Kasner points \cite {lim}. A
description of the dynamics of tilted spatially
homogeneous cosmologies of Bianchi type II (with a
perfect fluid with linear equation of state) has been presented
\cite {Hewitt}.
The Bianchi II cosmologies, while very special within
the whole Bianchi class, play a central
role since the Bianchi II state space is part of the
boundary of the state space for all higher Bianchi types
(including types VI$_0$  and VII$_0$). 
The class of tilted
Bianchi II cosmologies  can be described by a
set of expansion-normalized variables, and the state space is
bounded. (Note that expansion-normalized variables and $\beta$ -normalized variables 
are effectively the same close to the
initial singularity). In more detail \cite {Hewitt}, there is no equilibrium point
that is a local source, except in the
special case $\gamma =2$ (in which there is a
local source, namely a subset of the Jacobs disc). In the tilting situation there
are two Kasner circles, the standard Kasner
circle and the Kasner circle
with extreme tilt.  The evolution of tilted cosmologies of Bianchi type
II in the singular asymptotic regime is governed by
infinite heteroclinic sequences which contain orbits
that join two points between these two sets. 
This is analogous to the case of non-tilted spatially homogeneous
cosmologies of Bianchi types VIII and IX 
which exhibit so-called {\it Mixmaster oscillatory
behaviour} as the singularity is approached
into the past. Earlier work had shown that there
are no local sources in Bianchi type VI$_0$ models with a magnetic field,
so that in general these models also exhibit mixmaster behaviour to the past \cite{lkw}.

Consequently, the Bianchi
models have bifurcations at $\gamma = 2/3$ and $4/3$ (e.g., the
dimension of the unstable and stable manifolds of the
equilibrium points change).
There is no local source in general, and the models exhibit mixmaster 
behaviour to the past. A subset of models are past asymptotic to the flat 
Friedmann model
(i.e.,  the singularity is isotropic) if $\gamma > 4/3$. In particular, 
there is a bifurcation at $\gamma = \frac{2}{3}$ in both the tilting and 
magnetic field models in general relativity. We note that this is 
consistent with our bifurcation $\gam=4/3$ (with $\gam \rightarrow 2\gam$ 
in brane-world models).

\subsection{Extensions}
The earliest investigations of the initial singularity,
which used only isotropic fluids as a source of matter, suggested a 
matter-dominated isotropic 
singularity for all $\gam>1$ \cite{BDL,COLEY,chaos,isot}. 
However, it was shown in later work using anisotropic stresses \cite{BH}  
that the initial
singularity for a magnetic brane-world  could be either locally
isotropic or anisotropic. In particular,
 Barrow and Hervik \cite{BH} studied a class of Bianchi type I brane-world
models with a pure magnetic field and a perfect fluid with a linear barotropic $\gamma$-law equation of state.
They found that when $\gamma \ge \frac{4}{3}$,  the equilibrium point ${\cal F}_b$
is again a local source (past-attractor), but that there exists a second equilibrium point
denoted $PH_1$, which corresponds to a new brane-world solution with a non-trivial
magnetic field, which is also a local source. When $\gamma < \frac{4}{3}$, $PH_1$
is the only local source. This was generalized by \cite{hervik}, who
 presented a thorough investigation of the initial
singularity in brane-world cosmological models. 
It was shown that for a  class of spatially homogeneous brane-worlds with
anisotropic stresses, both local and nonlocal, the brane-worlds
could have either an isotropic singularity or an anisotropic singularity; indeed, using a continuity argument it was shown that  there exists a past
attractor for models with nonlocal anisotropic stresses of  type
${\cal P}_{\mu\nu}={\cal U} D_{\mu\nu}$ where $\ptl_t D_{\mu\nu}$ is sufficiently
small. Hence, there is a class of models with
${\cal P}_{\mu\nu}\neq 0$ which have an anisotropic past attractor.
How large this class is, and if this anisotropic past attractor
exists for generic brane-worlds, needs further  work. However, the analysis in the present paper is consistent with the results of \cite{BH,hervik} in which an isotropic singularity exists for $\gam>4/3$.\\

In general,  ${\cal P}_{\mu\nu}$ is not specified. 
It must be derived from the exact 5-dimensional field equations in a self-consistent way. In our analysis we have assumed that the effective nonlocal anisotropic stress is zero (in the fluid comoving frame). Indeed, this is the only assumption we have made. But it is expected that inclusion
of  ${\cal P}_{\mu\nu}$ will {\em not} affect the 
qualitative dynamical fatures of the models close to the initial singularity
(and we still expect isotropization at early times).
${\cal E}_{\mu\nu}$ can be irreducibly decomposed according to equation (\ref{1}).
Hence, we expect that ${\cal P}_{\mu\nu} \sim  {\cal U}  g_{\mu\nu}$ on dimensional
grounds, and so for a Friedmann brane close to the initial singularity
we expect that ${\cal P}_{\mu\nu} \sim a^2$ ${\cal U}C_{\mu\nu}$ (where $C_{\mu\nu}$
is slowly varying), which is consistent with the linear (gravitational) perturbation 
analysis (in a pure AdS bulk background) \cite{moko}. Hence, ${\cal P}_{\mu\nu}$ is
negligible dynamically close to the initial singularity.\\

The results might also be applicable in a number of more general situations. For example, in theories with field equations with higher-order curvature
 corrections (e.g., the four-dimensional brane world in the case of a Gauss-Bonnet term in the  bulk spacetime \cite{MT}), the results concerning stability are not expected to be affected since the curvature is negligible close to the initial singularity.

\section{Conclusions}

Therefore, the numerical analysis supports the fact  that in spatially
inhomogeneous $G_2$ brane-world cosmological models 
the  initial singularity is {\em isotropic\/} ~\cite{GW85}.\\

Therefore, unlike the situation in general relativity, it is plausible that typically the
initial singularity is isotropic in brane world cosmology.
Such a `quiescent' cosmology~\cite{Barrow}, in which the universe began in 
a highly regular state but subsequently evolved towards irregularity, 
might offer an explanation of why our Universe might have began its 
evolution in such a smooth manner and may provide a realisation of 
Penrose's ideas on gravitational entropy  and the second law of 
thermodynamics in cosmology~\cite{Penrose79}. More importantly, it is 
therefore possible that a quiescent cosmological period occuring in brane  
cosmology provides a physical scenario in which the universe starts off 
smooth and that naturally gives rise to the conditions for inflation to  
subsequently take place.

Cosmological observations indicate that we live in a Universe which is
remarkably uniform on very large scales. However, the spatial homogeneity and
isotropy of the Universe is difficult to explain within the
standard general relativistic framework since, in the presence of matter,
the class of solutions to the Einstein equations which evolve
towards a Robertson-Walker universe is essentially a set of measure
zero. In the
inflationary scenario, we live in
 an isotropic region of a potentially highly 
irregular universe as the result of an expansion phase in the early universe
thereby solving many of the problems of cosmology. Thus this
scenario can successfully generate a homogeneous and
isotropic Robertson-Walker-like universe from initial conditions which, in the
absence of inflation, would have resulted in a universe far
removed from the one we live in today. However, still only a restricted set
of initial data will lead to smooth enough conditions for the
onset of inflation.

Let us discuss this in a little more detail. Although inflation gives a natural solution of the horizon problem of the big-bang universe, inflation requires homogeneous initial conditions over the super-horizon scale, i.e.,
it itself requires certain improbable initial conditions.
When inflation begins to act, the universe must already be smooth on a scale of 
at least $10^5$ times the Planck scale. 
Therefore, we cannot say
that it is a solution of the horizon problem, though it reduces the problem by many orders 
of magnitude.  Many people have investigated how initial 
inhomogeneity affects the onset of inflation \cite{GP,KK}. Goldwirth and Piran \cite{GP}, who 
solved the full Einstein equations for a 
spherically symmetric  spacetime, found that
{\em small-field} inflation models
of the type of {\em new inflation} is so 
sensitive to initial inhomogeneity that it requires homogeneity over a region of several horizon 
sizes. 
{\em Large-field} inflation models such 
as {\em chaotic inflation} is not so affected by initial inhomogeneity but requires a 
sufficiently high average value of the scalar field over a region of several horizon
sizes \cite{Brandenberger}. Therefore,  including spatial inhomogeneities
accentuates the difference between models like new inflation and those
like chaotic inflation; inhomogeneities further reduce the measure
of initial conditions yielding new inflation, whereas the inhomogeneities
have sufficient time to redshift in chaotic inflation, letting the
zero mode of the field eventually drive successful inflation. In conclusion, although inflation is a possible causal
mechanism for homogenization and isotropization, there is a fundemental 
problem in that the initial conditions must be
sufficiently smooth in order for inflation to subsequently take
place \cite{COLEY}. We have found that  an isotropic singularity   in
brane world cosmology might provide for  the
 necessary sufficiently smooth initial
conditions to remedy this problem.

It would be of interest to study general inhomogeneous ($G_0$) brane 
world models. The exponential decay rates in the case $\gam>4/3$ are 
calculated in the Appendix (in the separable volume gauge using 
Hubble-normalized equations). The decay rates are essentially the same as 
in the $G_2$ case studied in section \ref{sectiondecayrate} (there are 
minor differences due to $\beta$-normalization and the absense of two 
tilt variables in the $G_2$ case: cf. equations (\ref{Sigcasy})-(\ref{Sigpasy})). This 
supports the possibility that in general brane world cosmologies have an 
isotropic singularity. We hope to further study $G_0$ brane world models numerically 
in the future.

\begin{acknowledgements}
AAC was funded by the Natural Sciences and Engineering Research Council
of Canada and YH was funded by a Killam Scholarship.

\end{acknowledgements}

\appendix

\section{Hubble-normalized $G_0$ equations}

We introduce $H$-normalized variables as follows:
\bea
	({\cal N}^{-1},\ E_\alpha{}^i) &:=& (N^{-1},\ e_\alpha{}^i)/H
\\
	(\dot{U}_\alpha,\ A^\alpha,\ \Sig^{\alpha\beta},\ N^{\alpha\beta},\
	R^\alpha) &:=& 
	(\dot{u}_\alpha,\ a^\alpha,\ \sigma^{\alpha\beta},\ 
	n^{\alpha\beta},\ \Omega^\alpha)/H
\\
	(\Omega,\ \Omega_u,\ Q_u) &:=&
	(\kappa^2\rho,\ \frac{6\kappa^2}{\lambda}\rho_u,\
	\frac{6\kappa^2}{\lambda} q_u) / (3 H^2)
\\
	{\cal H} &:=& \sqrt{\frac{6}{\lambda \kappa^2}} \sqrt{3} H\ .
\eea
Using the separable volume gauge: $\dot{U}_\alpha = r_\alpha$, with 
${\cal N}=1$, as the choice of temporal gauge,  and the Fermi-propagated 
gauge: $R^\alpha=0$, as the choice of spatial gauge, the equations are:

\bea
\lb{limdl13comts}
\ptl_{t}E_{\alpha}{}^{i}
& = & (q\,\d_{\alpha}{}^{\beta} -\Sig_{\alpha}{}^{\beta}
)\,E_{\beta}{}^{i} \\
\ptl_{t}{\cal H}
& = & -(q+1){\cal H}
\\
\lb{limdlrdot}
\ptl_t r_\alpha
& = & (q\,\d_\alpha{}^\beta - \Sig_\alpha{}^\beta)\,r_\beta+ 
\parb_{\alpha}q\\
\lb{limdladot}
\ptl_{t}A^{\alpha}
& = & (q\,\d^{\alpha}{}_{\beta} - \Sig^{\alpha}{}_{\beta}
)\,A^{\beta}
+ \frac{1}{2}\,\parb_{\beta}\Sig^{\alpha\beta}
\\
\lb{limdlsigdot}
\ptl_{t}\Sig^{\alpha\beta}
& = & (q-2)\,\Sig^{\alpha\beta} - 2N^{\langle\alpha}{}_{\gam}\,
N^{\beta\rangle\gam} + N_{\gam}{}^{\gam}\,N^{\langle\alpha\beta\rangle}
- \d^{\gam\langle\alpha}\,(\parb_{\gam}-r_{\gam})\,A^{\beta\rangle}
\nonumber \\
& & \hsp5 + \ \eps^{\gam\delta\langle\alpha}\,(\parb_{\gam}
-2A_{\gam})\,N^{\beta\rangle}{}_{\delta}
+ (\d^{\gam\langle\alpha}\,\parb_{\gam}
+A^{\langle\alpha})\,r^{\beta\rangle}
+ 3\Pi_{\rm tot}^{\alpha\beta}
\\
\lb{limdlndot}
\ptl_{t}N^{\alpha\beta}
& = & (q\,\d^{(\alpha}{}_{\delta}
+ 2\Sig^{(\alpha}{}_{\delta}
)\,N^{\beta)\delta}
-
\eps^{\gam\delta(\alpha}\,\parb_{\gam}\,\Sig^{\beta)}{}_{\delta} \\
\lb{limdlomdot}
\ptl_{t}\Om
& = & -\,\frac{\gam}{G_{+}}\,v^{\alpha}\,\parb_{\alpha}\Om
+ G_{+}^{-1}\,[\,2G_{+}q - (3\gam-2) - (2-\gam)\,v^{2}
- \gam\,(\Sig_{\alpha\beta}v^{\alpha}v^{\beta}) \nonumber \\
& & \hspace{55mm} - \ \gam\,(\parb_{\alpha}-2A_{\alpha})\,v^{\alpha}
+ \gam\,v^{\alpha}\,\parb_{\alpha}\ln G_{+}\,]\,\Om \\
\lb{limdlvdotf}
\ptl_{t}v^{\alpha}
& = & -\,v^{\beta}\,\parb_{\beta}v^{\alpha}
+ \d^{\alpha\beta}\,\parb_{\beta}\ln G_{+}
- \frac{(\gam-1)}{\gam}\,(1-v^{2})\,\d^{\alpha\beta}\,
(\parb_{\beta}\ln\Om-2r_{\beta}) \nonumber \\
& & + \ G_{-}^{-1}\,\Big[\,(\gam-1)\,(1-v^{2})\,(\parb_{\beta}
v^{\beta}) - (2-\gam)\,v^{\beta}\,\parb_{\beta}\ln G_{+}
\nonumber \\
& & \hspace{15mm}
+ \ \frac{(\gam-1)}{\gam}\,(2-\gam)\,(1-v^{2})\,v^{\beta}\,
(\parb_{\beta}\ln\Om-2r_{\beta})
+ (3\gam-4)\,(1-v^{2}) \nonumber \\
& & \hspace{15mm} + \ (2-\gam)\,(\Sig_{\beta\gam}v^{\beta}
v^{\gam}) + G_{-}\,(r_{\beta}v^{\beta})
+ [G_{+}-2(\gam-1)]\,(A_{\beta}v^{\beta})\,\Big]\,v^{\alpha}
\nonumber \\
& & - \ \Sig^{\alpha}{}_{\beta}\,v^{\beta}
- r^{\alpha} - v^{2}\,A^{\alpha}
+ \eps^{\alpha\beta\gam}\,N_{\beta\delta}\,v_{\gam}\,v^{\delta}
\\
\ptl_{t} \Omega_u
& = & 2(q-1)\Omega_u -\frac{4}{3+v^2} \Sig_{\alpha\beta} v^\alpha 
v^\beta \Omega_u + 2 A_\alpha Q_u^\alpha - \parb_\alpha Q_u^\alpha
+ v_\alpha X^\alpha
\\
\ptl_{t} Q_u^\alpha
& = & 2(q-1) Q_u^\alpha - \Sig^\alpha{}_\beta Q_u^\beta
+ 4 \frac{3 v^\alpha v^\beta - v^2 \delta^{\alpha\beta}}{3+v^2} A_\beta 
\Omega_u
\nonumber \\
& &
+\frac{2}{3+v^2} [ 2 v^\alpha v^\beta -(1+v^2)\delta^{\alpha\beta} ] 
r_\beta \Omega_u
- \parb_\beta (P_u \d^{\alpha\beta}+\Pi_u^{\alpha\beta})
+ X^\alpha
\eea
where
\bea
X^\alpha &=& \frac{\gam^2(1-v^2)}{6G_+{}^3G_-} \M \parb_\beta v^2
- \frac{\gam(1-v^2)^2}{6G_+{}^2G_-} \M \Om^2 {\cal H}^2 
	[ \parb_\beta \ln\Om - 2 r_\beta ] 
\nonumber\\
& &\quad
+ \frac{\gam^2(1-v^2)}{6G_+{}^2G_-} 
[\parb_\beta v^\beta +  (3-v^2) - \Sig_{\mu\nu} v^\mu v^\nu -2 A_\mu v^\mu ] 
v^\alpha \Om^2 {\cal H}^2
\\
\M & =& G_- \d^{\alpha\beta}+(\gam-1)v^\alpha v^\beta 
\\
q &=& 2\Sig^{2}+\frac{1}{2} (\Om_{\rm tot}+3 P_{\rm tot}) - \frac{1}{3}\,(\parb_{\alpha}
	-2A_{\alpha})\,r^{\alpha}
\\
r_\alpha &=& - \parb_\alpha \ln {\cal H}\ .
\eea
The constraints are:
\bea
\lb{limdl13comss}
0 & = &  2\,(\parb_{[\alpha}-r_{[\alpha}-A_{[\alpha})\,
E_{\beta]}{}^{i}
- \eps_{\alpha\beta\delta}\,N^{\delta\gam}\,E_{\gam}{}^{i} \\
\lb{limdlgauss}
0 & = & \ 1 - \Om_k - \Sig^{2} - \Om_{\rm tot} \\
\lb{limdlcodacci}
0 & = & \ \parb_{\beta}\Sig^{\alpha\beta}
+ (2\d^{\alpha}{}_{\beta}-\Sig^{\alpha}{}_{\beta})\,r^{\beta}
- 3A_{\beta}\,\Sig^{\alpha\beta}
- \eps^{\alpha\beta\gam}\,N_{\beta\delta}\,\Sig_{\gam}{}^{\delta}
+ 3 Q_{\rm tot}^\alpha
\eea
where
\be
\Om_k = -\,\frac{1}{3}\,(2\parb_{\alpha}
-2r_{\alpha}-3A_{\alpha})\,A^{\alpha}
+ \frac{1}{6}\,(N_{\alpha\beta}N^{\alpha\beta})
- \frac{1}{12}\,(N_{\alpha}{}^{\alpha})^{2}\ .
\ee
If we have (for $\gam>4/3$):
\bea
C_1^\star: & &\lim_{t\rightarrow -\infty} (E_\alpha{}^i, {\cal H}^{-1}, r_\alpha,
        A^\alpha, \Sig^{\alpha\beta}, N^{\alpha\beta}, \Om, v^\alpha, 
	\Om_u, Q_u^\alpha, \Om_b-1
        )
        = \mathbf{0}\ ,\nonumber
\\
C_2^\star: & &\ptl_i (E_\alpha{}^i, {\cal H}^{-1}, r_\alpha,
        A^\alpha, \Sig^{\alpha\beta}, N^{\alpha\beta}, \Om, v^\alpha,
        \Om_u, Q_u^\alpha, \Om_b-1
        )
        \quad\text{are bounded as \,\,\, $t \rightarrow -\infty$.} \nonumber
\\
C_3^\star: & & V= {\cal O}(f(t))\,\,\,\,\text{implies}\,\,\,\, \ptl_i V = {\cal O}(f(t))\,\,\,\,
        \text{ (asymptotic expansions in time} \nonumber
\\
        & & \text{can be differentiated with respect to the
        spatial coordinates).} \nonumber
\eea
then we can follow the analysis in \cite{lim0306118} to obtain the asymptotic decay rates.\\

\noindent\textbf{Stage 1}: First, $C_1^\star$ and $C_2^\star$ imply that $q \rightarrow 3\gam-1$.
Using $C_1^\star$, $C_2^\star$ and the evolution equations, we obtain from
Proposition 1 in \cite{lim0306118} in succession,
\bea
        E_\alpha{}^i &=& {\cal O}(\e^{[(3\gam-1)-\eps]t})
\\
\lb{limHrate}
        {\cal H}^{-1} &=&{\cal O}(\e^{[3\gam-\eps]t})
\\
        r_\alpha &=& {\cal O}(\e^{[(3\gam-1)-\eps]t})
\\
        A^\alpha &=& {\cal O}(\e^{[(3\gam-1)-\eps]t})
\\
        N^{\alpha\beta} &=& {\cal O}(\e^{[(3\gam-1)-\eps]t})\ .
\eea
It then follows from $C_1^\star$, $C_2^\star$, equation (\ref{limHrate}) and the algebraic expression for $\Om_b$ that
\be
        \Om =  {\cal O}(\e^{[3\gam-\eps]t})\ .
\ee

\noindent \textbf{Stage 2}: Using $C_1^\star$, $C_2^\star$, $C_3^\star$ and the evolution equations, we obtain from
Proposition 1 in succession,
\bea
        v^\alpha &=& {\cal O}(\e^{[(3\gam-4)-\eps]t})
\\
        \Sig^{\alpha\beta} &=& {\cal O}(\e^{[3(\gam-1)-\eps]t}
                                +\e^{[2(3\gam-4)-\eps]t})
\\
	Q_u^\alpha &=& {\cal O}(\e^{[(3\gam-4)-\eps]t}) 
\\
	\Om_u &=& {\cal O}(\e^{[2(3\gam-4)-\eps]t})\ .
\eea

\noindent \textbf{Stage 3}: First,
\be
        q = 3\gam-1 + {\cal O}(\e^{[2(3\gam-4)-\eps]t})\ .
\ee

As in \cite{lim0306118}, we use $C_1^\star$, $C_2^\star$, $C_3^\star$ and Proposition 4 and the evolution equations to obtain
\bea
        E_\alpha{}^i &=& \e^{(3\gam-1)t} [ \hat{E}_\alpha{}^i + {\cal O}(g) ]
\\
        {\cal H}^{-1} &=& \e^{3\gam t} [ \hat{{\cal H}}^{-1} + {\cal O}(v^2) ]
\\
        r_\alpha &=& \e^{(3\gam-1)t} [ \hat{r}_\alpha + {\cal O}(g) ]
\\
        A^\alpha &=& \e^{(3\gam-1)t} [ \hat{A}^\alpha + {\cal O}(g) ]
\\
        N^{\alpha\beta} &=& \e^{3(\gam-1)t} [ \hat{N}^{\alpha\beta} 
							+ {\cal O}(g) ]
\\
        \Om &=& \e^{3\gam t} [ \sqrt{12} \hat{{\cal H}}^{-1}  + {\cal O}(v^2) ]\ 
.
\\
	v^\alpha &=& \hat{v}^\alpha \e^{(3\gam-4)t} + {\cal O}(h)
\eea

\be
	\Sig^{\alpha\beta} =\left\{ \begin{array}{ll}
	\frac{6\gam}{3\gam-5} \hat{v}^{\langle \alpha} \hat{v}^{\beta \rangle}
	\e^{2(3\gam-4)t} + \hat{\Sig}^{\alpha\beta} \e^{3(\gam-1)t}
	+ {\cal O}(vh)
	& \frac{4}{3} < \gam < 2,\ \gam\neq \frac{5}{3}
	\\
	10 \hat{v}^{\langle \alpha} \hat{v}^{\beta \rangle} t\e^{2t}
	+ \hat{\Sig}^{\alpha\beta} \e^{2t}
	+ {\cal O}(vh)
	& \gam = \frac{5}{3}
	\end{array} \right.
\ee

\bea
	Q_u^\alpha &=& -2\gam \hat{v}^\alpha \e^{(3\gam-4)t}
		+ \hat{Q}_u^\alpha \e^{2(3\gam-2)t}
		+ {\cal O}( h )
\\
	\Om_u &=& -\frac{3}{2} \gam^2 \hat{v}^2 \e^{2(3\gam-4)t}
		+ \hat{\Om}_u \e^{2(3\gam-2)t}
		+ {\cal O}( vh )   
\eea
where $g={\cal O}(\e^{[3(\gam-1)-\eps]t}+\e^{[2(3\gam-4)-\eps]t})$, 
the dominant terms being $\Sig^{\alpha\beta}$ and $v^2$;
and $h = {\cal O}(\e^{3(\gam-1)t} + \e^{[3(3\gam-4)-\eps]t})$.

Finally we eliminate the epsilons from $g$, $h$ and other ${\cal O}$ terms by 
repeating (and using the results from) Stage 3:
\bea
        g = {\cal O}(\Sig) &=&\left \{ \begin{array}{ll}
                {\cal O}(\e^{3(\gam-1)t}+\e^{2(3\gam-4)t})
                & \text{$\frac{4}{3} < \gamma < 2, \gamma\neq \frac{5}{3}$}
                \\
                {\cal O}(t \e^{2t})
                & \text{ $\gamma = \frac{5}{3}$}
                \end{array} \right.
\\
        h &=& {\cal O}(r+v^3) = {\cal O}(\e^{(3\gam-1)t}
                                + \e^{3(3\gam-4)t})
\\
	q &=& 3\gam-1 + {\cal O}(\e^{2(3\gam-4)t})
\eea

\textbf{Summary}: as $t \rightarrow -\infty$ we have that
\bea
        E_\alpha{}^i &=& \e^{(3\gam-1)t} [ \hat{E}_\alpha{}^i + {\cal O}(g) ]
\\
        {\cal H}^{-1} &=& \e^{3\gam t} [ \hat{{\cal H}}^{-1} + {\cal O}(v^2) ]
\\
        r_\alpha &=& \e^{(3\gam-1)t} [ \hat{r}_\alpha + {\cal O}(g) ]
\\
        A^\alpha &=& \e^{(3\gam-1)t} [ \hat{A}^\alpha + {\cal O}(g) ]
\\      
        N^{\alpha\beta} &=& \e^{3(\gam-1)t} [ \hat{N}^{\alpha\beta}
                                                        + {\cal O}(g) ]
\\                      
        \Sig^{\alpha\beta} &=&\left \{ \begin{array}{ll}
        \frac{6\gam}{3\gam-5} \hat{v}^{\langle \alpha} \hat{v}^{\beta \rangle}
        \e^{2(3\gam-4)t} + \hat{\Sig}^{\alpha\beta} \e^{3(\gam-1)t}
        + {\cal O}(vh)
        & \frac{4}{3} < \gam < 2,\ \gam\neq \frac{5}{3}
        \\
         10 \hat{v}^{\langle \alpha} \hat{v}^{\beta \rangle} t\e^{2t}
        + \hat{\Sig}^{\alpha\beta} \e^{2t}
        + {\cal O}(vh)
        & \gam = \frac{5}{3}
        \end{array} \right.
\\
        \Om &=& \e^{3\gam t} [ \sqrt{12} \hat{\cal H}^{-1}  + {\cal O}(v^2) ]
\\
        v^\alpha &=& \hat{v}^\alpha \e^{(3\gam-4)t} + {\cal O}(h)
\\
        Q_u^\alpha &=& -2\gam \hat{v}^\alpha \e^{(3\gam-4)t}  
                + \hat{Q}_u^\alpha \e^{2(3\gam-2)t}
                + {\cal O}( h )
\\
        \Om_u &=& -\frac{3}{2} \gam^2 \hat{v}^2 \e^{2(3\gam-4)t}
                + \hat{\Om}_u \e^{2(3\gam-2)t}
               + {\cal O}( vh )\ ,
\eea
where
\bea
        g &=& {\cal O}(\Sig) = \left \{ \begin{array}{ll}
                {\cal O}(\e^{3(\gam-1)t}+\e^{2(3\gam-4)t})
                & \text{$\frac{4}{3} < \gamma < 2, \gamma\neq \frac{5}{3}$}
                \\
                {\cal O}(\tau \e^{2t})
                & \text{ $\gamma = \frac{5}{3}$}
                \end{array}  \right.
\\ 
        h &=& {\cal O}(r+v^3) = {\cal O}(\e^{(3\gam-1)t}
                                + \e^{3(3\gam-4)t})\ .
\eea

\end{document}